\documentclass[12pt]{article}
\usepackage[utf8]{inputenc}
\usepackage{graphicx}
\usepackage{footmisc}
\usepackage{subcaption} 
\usepackage{url}
\usepackage{multirow}
\usepackage[noadjust]{cite}
\usepackage{amsmath,amsthm}
\usepackage{amssymb,amsfonts}
\usepackage{booktabs}
\usepackage{makecell}
\usepackage{color}
\usepackage{ccaption}
\usepackage{float}
\usepackage{fancyhdr}
\usepackage{caption}
\usepackage{xcolor,stfloats}
\usepackage{comment}
\graphicspath{{figures/}}
\usepackage{cuted}  
\usepackage{bm}  
\usepackage{algorithm}
\usepackage{algorithmic}

\usepackage[figuresright]{rotating}
\usepackage[left=1in, right=1in, top=1in, bottom=1in]{geometry}
\newcommand{\upcite}[1]{\textsuperscript{\cite{#1}}}
\title{Role Identification based Method for Cyberbullying Analysis in Social Edge Computing}
\author{Runyu Wang\footnotemark[1] , Tun Lu\footnotemark[1] \footnotemark[2] , Peng Zhang\footnotemark[1] \footnotemark[2] , and Ning Gu\footnotemark[1]}
\date{}

\begin{document}

\maketitle
\renewcommand{\thefootnote}{\fnsymbol{footnote}} 
\footnotetext[1]{Runyu Wang, Tun Lu, Peng Zhang, and Ning Gu are with Fudan University, Shanghai 200433, China. 
E-mail:  20110240082@fudan.edu.cn; lutun@fudan.edu.cn; zhangpeng\_@fudan.edu.cn; ninggu@fudan.edu.cn.} 
\footnotetext[2]{To whom correspondence should be addressed.} 

\begin{abstract}
Over the past few years, many efforts have been dedicated to studying cyberbullying in social edge computing devices, and most of them focus on three roles: victims, perpetrators, and bystanders. If we want to obtain a deep insight into the formation, evolution, and intervention of cyberbullying in devices at the edge of the Internet, it is necessary to explore more fine-grained roles. This paper presents a multi-level method for role feature modeling and proposes a differential evolution-assisted K-means (DEK) method to identify diverse roles. Our work aims to provide a role identification scheme for cyberbullying scenarios for social edge computing environments to alleviate the general safety issues that cyberbullying brings. The experiments on ten real-world datasets obtained from Weibo and five public datasets show that the proposed DEK outperforms the existing approaches on the method level. After clustering, we obtained nine roles and analyzed the characteristics of each role and their evolution trends under different cyberbullying scenarios. Our work in this paper can be placed in devices at the edge of the Internet, leading to better real-time identification performance and adapting to the broad geographic location and high mobility of mobile devices.
\end{abstract}

\section{Introduction}
\label{s:introduction}
\noindent
As a decentralized computing approach, edge computing focuses on processing and analyzing data closer to the data source rather than relying on a central cloud data center\upcite{1}. In the Internet of Things, edge computing can efficiently process data from countless devices, enabling real-time data processing, faster decision-making, and higher operational efficiency\upcite{2}. As a new paradigm of edge computing, social edge computing enters the field of view of scholars, which fully uses the ubiquitous, individually owned edge devices and the collective computing power, intelligence and data of device owners\upcite{3}. In addition, device owners involved in edge artificial intelligence (AI) applications are referred to as edge crowd\upcite{3}.

With the rapid popularization of the Internet, online applications in various social edge computing devices have become an essential source of social hot spots and an important channel for people to express their views. In this context, the number of device owners involved in various AI applications continues proliferating. However, the proliferation of the edge crowd and the rapid iteration of social edge computing devices has led to the evolution of a new form of bullying: cyberbullying.

Rathore et al.\upcite{4} defined cyberbullying as the phenomenon of intentionally and repeatedly harassing or harming someone online. Many victims of cyberbullying feel disconnected from their school and community, suffer traumatic stress, and lose interest in the things they once enjoyed\upcite{5}. Shi et al.\upcite{6} stated that cyberbullying depicts the harmful effects of posting or sending personal information in the digital space, which can cause embarrassment and humiliation to the victim. However, most previous studies about cyberbullying focus on its detection, while the identification of diverse roles under different cyberbullying scenarios is ignored. Inspired by social roles in Refs. \cite{7,8}, cyberbullying roles in this paper are defined as members who share similar behavioral patterns when participating in cyberbullying in edge AI applications on social edge computing devices. Previous studies mainly investigated three roles, i.e., perpetrator, victim, and bystander\upcite{9}. Although the three-role-based framework can be used to draw relevant conclusions on cyberbullying, this framework is too coarse-grained to observe cyberbullying behavior accurately. Specifically, the identification of multiple different roles will go beyond the traditional “perpetrator-victim-bystander” framework and expand the framework to include more granular roles, which will facilitate the simulation of a more realistic cyberbullying environment and the targeted intervention of different roles on social edge computing devices. In addition, the dynamic change of fine-grained roles over time is also lacking in current research. The interaction between different roles and the interaction between roles and the environment can lead to the emergence of group phenomena, which is also conducive to exploring the causes of macro phenomena from the micro level. 

To sum up, to have a more in-depth insight into the formation, evolution, and intervention of cyberbullying, it is necessary to explore more fine-grained roles. It is worth mentioning that the individuals or user groups involved in the cyberbullying mentioned in this paper are considered to be the edge crowd. Besides, inspired by Ref. \cite{3}, the social edge computing devices we study can be used for Sina Weibo, one of China’s biggest social media platforms. Hence, this work aims to study the following research questions:

$\bullet$ \textbf{RQ1:} What kinds of roles are there when cyberbullying occurs in edge AI applications on social edge computing devices? What are the typical behavioral characteristics of these roles?

$\bullet$ \textbf{RQ2:} What is the difference in member size in different roles under different cyberbullying scenarios in edge AI applications on social edge computing devices?

$\bullet$ \textbf{RQ3:} What are the typical factors influencing the distribution of the number of roles under cyberbullying scenarios on social edge computing devices?

To explore the above issues in edge AI applications on social edge computing devices, we first propose a fine-grained and multi-level modeling method for role features. Then, based on the characteristics of the continuous and categorical variables in the corpus used in this work, we propose a differential evolution-assisted K-means (DEK) for role identification and conduct numerical studies based on five public datasets and ten real-world datasets. Finally, we identify nine cyberbullying roles and analyze the corresponding behavioral characteristics. Moreover, we explore the distribution of members occupied in different roles over time and their influencing factors.

In this paper, we innovatively utilize the behavioral capacity and emotional characteristics of the edge crowd who engage in cyberbullying on social edge computing devices, which fills a gap in the previous machine-centric edge computing paradigm. Our proposed approach effectively divides the edge crowd involved in cyberbullying on social edge computing devices into different roles, thus helping to explore the deep knowledge of cyberbullying on social edge computing devices. In addition, our work inspires the development of new applications to address the societal challenge of cyberbullying in edge AI applications on social edge computing devices. In summary, our work demonstrates the following contributions:

\textbf{1) Dataset:} We collect a cyberbullying dataset from social edge computing devices, including an edge crowd of 3,231 users and more than 1.47 million text posts related to ten cyberbullying scenarios. This dataset contains diverse topics of widespread concern and combines the presence or absence of four criteria to facilitate future research on cyberbullying.

\textbf{2) Feature engineering:} For edge crowd, we propose a fine-grained and multi-level role feature modeling method, namely content-based, sentiment-based, and user-based feature modeling, which helps to deeply understand the behavioral characteristics of edge crowd and significantly improve the performance of role identification.

\textbf{3) Algorithm:} We propose a clustering algorithm named DEK in the mixed categorical-continuous space composed of features of edge crowd. In DEK, the improved differential evolution is utilized to avoid local optimal and to handle mixed categorical-continuous variables during the clustering process.

\textbf{4) Analysis:} To the best of our knowledge, we are the first to conduct fine-grained and multi-level role identification on social edge computing devices for different cyberbullying scenarios. We further analyze the member size and the influencing distribution factors of these identified roles under different scenarios on social edge computing devices.

The remainder of this paper is organized as follows: Section 2 summarizes the related work on cyberbullying detection and social role identification. Sections 3 and 4 illustrate the details of the data sources and feature engineering, respectively. The proposed DEK and the numerical study are elaborated in Section 5. Section 6 describes the qualitative analysis of the identified roles. Finally, we conclude our study and present future work in Section 7.

\section{Related Work}
\noindent
Cyberbullying detection is the basis of cyberbullying role identification. In the edge AI applications on social edge computing devices, different members participate in different activities, whether implementing cyberbullying is an important part of the role division. Whether or not cyberbullying has been committed requires machine learning or other methods to detect published posts. In addition, role identification in other areas of edge AI applications also has some similarities with cyberbullying. Therefore, in this section, we review the progress related to our work according to the following two categories: cyberbullying detection and social role identification.

\subsection{Cyberbullying detection}
\noindent
Current studies on cyberbullying in social computing mainly focus on its detection. Cyberbullying detection is usually a classification problem in which appropriate features are extracted from a data sample (e.g., posts, messages, or comments from social media platforms) and put into the classifier to judge whether the data sample belongs to bullying data\upcite{10}. In Ref. \cite{11}, Al-Garadi et al. reviewed the related research on detecting cyberbullying messages. This literature also highlighted some problems in cyberbullying and proposed new research directions for scholars. According to the implementation and characteristics of cyberbullying detection methods, we divide them into rule-based, machine learning-based, and deep learning-based detection methods.

The rules-based method matches text with predefined rules to identify cyberbullying. For example, Chen et al.\upcite{12} proposed the Lexical Syntactic Feature architecture to detect offensive content and identify potential aggressive users in social media. The authors introduced hand-crafted syntactic rules to identify abusive harassment. Rule-based methods have previously been shown to identify cyberbullying texts. However, such methods are heavily limited to predefined rules and need to be more adaptable to constantly updated and complex cases.

Currently, researchers use machine learning methods including, but not limited to, support vector machine (SVM)\upcite{13,14}, logistic regression (LR)\upcite{15}, decision tree (DT)\upcite{16}, naive Bayes (NB)\upcite{17}, random forest (RF)\upcite{18}, reinforcement learning\upcite{19}, etc., to detect cyberbullying. For example, Davidson\upcite{20} used a crowdsourced hate speech dictionary to collect a Twitter dataset and extracted multiple text-based features of tweets. After testing with five machine learning algorithms (LR, NB, DT, RF, and linear SVM), the authors found that LR and linear SVM classifiers outperformed other algorithms. Liu et al.\upcite{21} studied the feasibility of improving cyberbullying detection on social networks through machine learning. They found that LR and SVM were superior to NB and DT in cyberbullying detection tasks. Balakrishnan et al.\upcite{22} improved the detection performance of NB, RF, and J48 DT by using empirical evidence of user personality, which the authors combined with other relevant characteristics, such as emotion and mood, for cyberbullying detection. 

Deep learning uses deep neural network models to process data such as text and images and does not require an explicit feature extraction strategy. More and more researchers are applying deep learning models to detect cyberbullying on social networks, including but not limited to the use of convolutional neural networks (CNN)\upcite{23}, recurrent neural networks (RNN)\upcite{24}, long short-term memory (LSTM)\upcite{24}, bi-directional LSTM (Bi-LSTM)\upcite{25}, bi-directional gated recurrent unit (Bi-GRU)\upcite{26}, attention mechanisms\upcite{27,28}, graph neural network (GNN)\upcite{29,30}, etc. Lu et al.\upcite{23} proposed a character-level CNN to identify whether the text in social media contains cyberbullying. The authors used characters as the smallest learning unit, allowing the model to overcome spelling errors and deliberate confusion in real-world corpora. Sultan et al.\upcite{24} proposed a new LSTM-CNN method to detect cyberbullying in online text content. Using the strengths of LSTM and CNN in capturing long-term dependencies and extracting relevant features from text data, the authors demonstrated the effectiveness of the LSTM-CNN approach in accurately identifying cyberbullying content. Chen et al.\upcite{25} proposed a hybrid cyberbullying speech detection model based on deep Bi-LSTM. This model improved the accuracy of Chinese cyberbullying detection through deep Bi-LSTM bidirectional coding. Kumar et al.\upcite{26} proposed a hybrid model, which learns sequential semantic representation and spatial location information using a Bi-GRU with self-attention. Cheng et al.\upcite{27} proposed an unsupervised cyberbullying detection framework based on temporal information, mainly consisting of representation and multi-task learning modules. The representation learning module used hierarchical attention to model text features. Yi et al.\upcite{28} proposed a cross-platform cyberbullying detection framework in which the self-attention mechanism-based Transformer model can define a potential feature space to reconcile the differences between the source and target platforms. Maity et al.\upcite{29} proposed a GNN-based multi-tasking framework, which solved the emotion-assisted cyberbullying detection problem in code-mixed languages. Li et al.\upcite{30} introduced an innovative framework based on the graph isomorphism network, which integrated the multi-modal feature construction method and multi-feature combination voting strategy to detect ironic cyberbullying.

\subsection{Social role identification}
\noindent
Research has been done on role identification in edge AI applications on social edge computing devices in specific fields. Social roles are often defined as a set of activities individuals engage in and the range of behaviors expected within a group\upcite{31}. Gleave et al.\upcite{32} stated that the best way to identify social roles is to start from the level of meaningful social actions, including identifying relevant behavioral laws and unique positions in social networks, as well as identifying abstract theoretical categories. In previous studies on social role identification in online communities, Wijenayake et al.\upcite{33} summarized the common role identification methods into manual coding, qualitative analysis, discourse analysis, cluster analysis, hierarchical division, etc.

In the non-cyberbullying domain, Yang et al.\upcite{7} introduced a general framework to define emergent roles in online communities. They manipulated a set of behavioral features to represent each component and used Gaussian Mixture Model (GMM) to extract the functional roles occupied by participants. To understand the role distribution in micro-lending platforms and help the team to operate more effectively, Sun et al.\upcite{8} conducted three-level role modeling on members’ lending, social network, and communication behaviors of the peer-to-peer micro-funding platform Kiva.org. They identified social roles in Kiva.org using GMM and used regression analysis to show how the distribution of social roles in a team affects the amount a team lent. Mukta et al.\upcite{34} proposed a weighted hybrid machine learning model to predict role identities from users’ social network interactions. Sun et al.\upcite{35} collected a large corpus of greeting card messages covering three different occasions. They found a wide range of gender stereotypes in this corpus through topic modeling, odds ratio and Word Embedding Association Test. The authors suggested that greeting card messages sent to women tend to be about their appearance and family. In contrast, those sent to men tend to be about work achievements, and greeting card messages sent to older people are less gender-related than those sent to other age groups. Their survey showed that most people want to be informed about gender roles in greeting card messages and avoid potential gender stereotypes. However, they do not want machines to interfere too much or modify their messages. Gondal\upcite{36} used loglinear models and a data mining technique called “association rules” to analyze data from the 2004 social networks component of the General Social Survey. This study explored how different relationship roles (such as parents, siblings, co-workers, and neighbors) and focal points (workplace, community, and group members) change the meaning of relationships. By analyzing publicly available LinkedIn recommendations, Erfani et al.\upcite{37} proposed a data-driven approach to examine female architecture leaders’ likability and competence dilemma. The results showed that female leaders in the construction industry were seen as just as competent as their male counterparts but far less likable. The authors contributed to social role theory and role consistency theory by highlighting differences in gender roles and leadership roles for women in the construction industry. Yang et al.\upcite{38} presented a two-stage role identification method that iteratively optimizes the role assignments of team members to predict team quality best. The proposed role identification model can not only accurately predict the quality of teamwork but also provide explicable results of student role assignments.

In the cyberbullying domain, Balakrishnan et al.\upcite{39} proposed a detection model to identify bullying patterns in the Twitter community based on the relationship between personality traits and cyberbullying, which was determined by the Big Five personality model and the Dark Triad model. Based on 9,484 tweets, the authors used RF to classify users into four roles: bully, attacker, spammer, and normal. Jacobs et al.\upcite{40} studied fine-grained role detection of cyberbullying in English and Dutch in a social media corpus to distinguish between bullies, victims, and bystanders. The authors proposed two different experimental settings for optimizing and comparing linear task-specific classification algorithms and fine-tuning the pre-trained Transformer model. These two experimental settings proved that participant roles could be classified, and satisfactory results were obtained. In Ref. \cite{41}, the authors identified profanity words from Ask.Fm corpus and analyzed the distribution of profanity words in four different roles: harasser, victim, bystander who assisted the bully and bystander who protected the victim in cyberbullying based on dictionaries.

\subsection{Summary}
\noindent
The above is the existing research work on cyberbullying detection and social role identification. Although many scholars have proposed different methods for role identification, there is still a lack of role identification methods for cyberbullying.

\section{Research Site: Sina Weibo}
\noindent
We obtain users’ public information and comments from Sina Weibo in this work. We specifically select users or topics that attracted colossal attention to form ten scenarios as our work’s real-world datasets. The real-world datasets reflect different topics, official intervention strategies, victim’s subjective faults, and public perceptions of victims. The real-world datasets include 3,231 users and over 1.47 million textual comments. The ten scenarios are as follows:

\textbf{1) The Tie Li Case (Scenario 1):} Tie Li resigned as the head coach of the China national football team in December 2021. On February 1, 2022, China suffered an emphatic 3-1 defeat at the hands of Vietnam. Although Li has left the national team, he was still subjected to online violence. 

\textbf{2) The Alibaba Sexual Assault Case (Scenario 2):} In August 2021, Zhou, an Alibaba employee, claimed she had been sexually abused by her colleagues Wang and Zhang. Then, Wang’s wife and Zhang’s wife refuted it but were attacked online. After one month, the local police issued a briefing that Zhou had exaggerated the truth. 

\textbf{3) The Shengbin Lin Case (Scenario 3):} In 2017, Shengbin Lin’s wife and children died in a fire caused by his nanny. In August 2021, Lin’s announcement of his remarriage and subsequent childbirth raised a barrage of questions and abuse. 

\textbf{4) The Jinglei Lee Divorce Case (Scenario 4):} On December 17, 2021, Jinglei Lee posted accusations on Weibo against Leehom Wang, claiming Wang’s multiple infidelities. While Lee has received public support and sympathy, many hostile voices have been against her. 

\textbf{5) The Zheng Yu’s Cat Case (Scenario 5):} In November 2021, a scene of a poisoned cat dying in Zheng Yu’s TV drama sparked concerns. Although the crew has repeatedly denied that they abused the cat, public doubts persisted. 

\textbf{6) The Ge Jiang’s Mother Case (Scenario 6):} Ge Jiang’s mother demanded an investigation into her finances and trademark registration, claiming this move was to protect her dead daughter. Some netizens made accusations against her. 

\textbf{7) The Account Selling Case (Scenario 7):} China Eastern Airlines flight MU5375 crashed in March 2022. After one month, a victim’s ex-girlfriend fraudulently obtained the victim’s game account and sought profits, which was met with great condemnation. 

\textbf{8) The Confirmed COVID-19 Patient Case (Scenario 8):} On December 8, 2021, the activity tracking and private information of Zhao, a confirmed COVID-19 patient in Chengdu, was exposed. Some netizens attacked Zhao’s personal life. 

\textbf{9) The Jing Wu’s IP Address Case (Scenario 9):} In May 2022, Jing Wu, a well-known actor, was abused by some brainless netizens because his IP address was in Thailand. In fact, he was filming The Megalodon 2. in Thailand. 

\textbf{10) The Godfrey Gao’s Death Case (Scenario 10):} On November 27, 2019, Godfrey Gao collapsed during the filming of a Zhejiang TV show and died after rescue. Many netizens set off a burst of accusations on Zhejiang TV.

The above scenarios involve diverse topics, i.e., sports (Scenario 1), entertainment (Scenarios 4, 5, 9, and 10), marriage (Scenarios 2, 3, and 4), ethics \& morality (Scenarios 6 and 7), pets (Scenario 5), COVID-19 (Scenarios 8 and 9) and gender (Scenario 2). In addition, the official intervention appears in Scenarios 2 and 4 but not in other scenarios. The victim has a subjective fault in Scenario 8, and the public knows the victim’s experience in Scenarios 3, 6, and 9, while the victim’s subjective fault and experience are unknown in other scenarios. 

\section{Feature Engineering}
\noindent
In previous works, Capua et al.\upcite{42} proposed a method to automatically detect bullying traces on social networks, which considered syntactic, semantic, emotional, and social features. Bozyiit et al.\upcite{43} analyzed and concluded that some features, such as forwarding and sender location, are closely related to cyberbullying in social media. They suggested that the classifier’s performance is improved when the relevant social media attribute features are used together with text mining methods. First, the data collected from social edge computing devices are text-based posts, so the characteristics of text-based content are the most important source of information. Second, the emotional changes contained in the posts are often related to the occurrence of cyberbullying incidents. Third, when cyberbullying occurs, the basic user characteristics of the edge crowd often play an important role in distinguishing cyberbullying from non-cyberbullying instances. Therefore, to identify the social roles people enact when engaging in cyberbullying in edge AI applications on social edge computing devices, we present a fine-grained and multi-level modeling method for role features composed of three levels: content-based, sentiment-based, and user-based. These features are used as variables at each level to construct clustering models and generate reasonable groups of the edge crowd as potential social roles.

\subsection{Content-based features}
\noindent
Content-based features are the most frequently used and accessible for detecting cyberbullying. Generally, some recognized dominant features, such as malicious words or keywords with significant offensiveness, can be selected as content features. Then, language models convert these content features into a form that machine learning algorithms can accept.

\textbf{1) Core insulting \& extended insulting words}

Since some cyberbullying messages contain insulting words, these words are a good indication of the happening of bullying\upcite{44,45}. In this work, insulting words in Noswearing\upcite{46}, an external language resource website, are utilized to form a core insulting word set. However, due to the rapid evolution of profanity slang and the complexity and synonymy of the language, the core insulting word set cannot cover all possible insulting words\upcite{45,47}. To overcome this limitation, Zhao et al.\upcite{45} used the word embedding method to expand and weigh insulting words.

Therefore, we define the core insulting words as the insulting seeds and subsequently utilize the Tencent Embedding Corpora\upcite{48} to expand the insulting seeds to form an extended insulting word set. By comparing the words in the above two insulting word sets with the words in the real-world datasets, we obtain the number of core and extended insulting words.

\textbf{2) Keywords}

Some studies take keywords that reflect the text’s theme into consideration of feature engineering in cyberbullying detection. Dadvar et al.\upcite{49} counted the number of person pronouns according to a pronoun list and considered issues such as race, religion, and physical appearance to detect cyberbullying. Perera et al.\upcite{15} compiled a set of words associated with racism, sex, and swear to detect cyberbullying. 

In this work, the following features are included: personal pronouns, racial words, and specific keywords involving gender and physical characteristics.

\subsection{Sentiment-based features}
\noindent
Sentiment-based features refer to the emotional information in the posts by the edge crowd on social edge computing devices. In general, a cyberbullying text has a certain emotional tendency. Therefore, consideration is given to the sentiment polarity, emotional type, and sentiment score.

\textbf{1) Sentiment polarity}

Sentiment polarity is usually categorized as positive, negative, and neutral, indicating how users feel when they post online. Text sentiment analysis can provide useful features for identifying harmful or abusive content\upcite{50}. According to Ref. \cite{51}, sentiment is a feeling triggered by words, images, or videos. The positive, neutral, or negative sentiment score can be used as a feature of cyberbullying detection. 

In this work, we utilize the Baidu AI Cloud\upcite{52} to identify sentiment polarity. First, we upload the positive and negative sentiment corpus of specific application scenarios, more than 100,000 Sina Weibo corpus with sentiment annotation, to train the user-customized sentiment tendency model through the customization API. Next, we use the trained user-customized sentiment tendency model to predict the sentiment polarity, including positive, neutral, and negative for texts.

\textbf{2) Emotional type}

Singh et al.\upcite{44} stated that words describing anger, negative emotions, sadness, and anxiety indicate cyberbullying. Nahar et al.\upcite{51} proposed a graph-based approach to detecting perpetrators and victims in the collected dataset, where emotion was selected as the desired feature. For cyberbullying detection, Foong et al.\upcite{53} used Linguistic Inquiry and Word Count to capture negative emotions and words expressing sadness, anxiety, and anger. 

Therefore, we utilize DLUT-Emotionontology\upcite{54} to divide the emotion of texts into seven types: happy, good, surprise, anger, sad, fear, and disgust.

\textbf{3) Sentiment score}

Jacob et al.\upcite{55} judged whether a text is cyberbullying according to the score generated in the emotion extraction stage and the pre-defined threshold. Based on the sentiment dictionary, Davidson et al.\upcite{20} assigned a sentiment score to each tweet, which was used as a classification feature for language detection. Lee et al.\upcite{56} deployed a model to collect user and text information from Twitter and developed an automatic cyberbullying detection model based on text, mood score, and other users’ information.

In this work, we utilize DLUT-Emotionontology to score the emotion of the text.

\subsection{User-based features}
\noindent
There is a relationship between the users’ social profiles and the occurrence of cyberbullying\upcite{57}. The role of cyberbullying has specific demographic characteristics and standpoints on social edge computing devices, representing its behavioral context and views.

\textbf{1) Basic user information}

Raphel et al.\upcite{58} stated that gender significantly impacts the discrimination between victims and perpetrators. Some studies showed that anonymity has a positive impact on cyberbullying\upcite{59}. In addition, perpetrators can disguise themselves by hiding their true identities\upcite{60}. Cheng et al.\upcite{61} established a cyberbullying detection model that extracted five features: user, image, profile, post timestamp, and text information.  

From the personal public information of the edge crowd, we select their gender, account level, authentication, the number of historical posts, followers, and fans as the users’ basic features.

\textbf{2) User activities}

User activity features include the number of historical total posts, liked or disliked posts, and hashtag activity\upcite{47}. These features can measure users’ online communication activities and play an essential role in predicting cyberbullying\upcite{62,63}. Rafiq et al.\upcite{64} proposed a multi-stage cyberbullying detection solution using user features, including total individual comment polarity and subjectivity, total negative words, and total negative comments.

In this work, we analyze the sentiment polarity of all posts on users’ homepages and count the proportion of negative polarity posts. Additionally, we treat the number of likes and comments when cyberbullying occurs and the number of likes, reposts, and comments of all user historical posts as user-activity-based features.

\section{Method for Role Identification} 
\noindent
After the fine-grained modeling of the edge crowd at three levels in Section 4, we develop a K-means clustering algorithm assisted by differential evolution to obtain clusters of different members participating in cyberbullying on social edge computing devices. Then, we analyze clustering members who show similar behavior patterns to determine the role division in Section 6.

\subsection{K-means and differential evolution}
\noindent
K-means is a widely used clustering algorithm whose simplicity and speed are very appealing in practice\upcite{65}. The K-means algorithm is based on the Euclidean distance, which believes that the closer the distance between two points, the greater the similarity.

Suppose a dataset $\bm{D}=\left\{ \bm{x}_{1},\bm{x}_{2},\ldots,\bm{x}_{n}\right\}$ is divided into $K$ clusters $\bm{G}_{1},\bm{G}_{2},\ldots,\bm{G}_{K}$, $\bm{c}_{j}$ is the centroid of $\bm{G}_{j}$ ($j = 1,2,\ldots,K$), and the set of centroid points is denoted as $\bm{C}=\left\{\bm{c}_{1},\bm{c}_{2},\ldots,\bm{c}_{K} \right\}$. 
Each point in $\bm{G}_{j}$ is a vector that satisfies the following conditions:
\begin{itemize}
\item[$\bullet$]$\bm{G}_{i} \neq \emptyset, i = 1,2,\ldots,K$
\end{itemize}
\begin{itemize}
\item[$\bullet$]$\bm{G}_{i} \cap \bm{G}_{j} = \emptyset, i,j = 1,2,\ldots,K, i \neq j$
\end{itemize}
\begin{itemize}
\item[$\bullet$]${\textstyle \bigcup_{i = 1}^{K}{G}_{i}} =\left\{ \bm{x}_{1},\bm{x}_{2},\ldots,\bm{x}_{n}\right\}$
\end{itemize}

The K-means clustering algorithm has the following steps:

\textit{\textbf{1)}} Randomly select $K$ points $\bm{c}_{1},\bm{c}_{2},\ldots,\bm{c}_{K}$ as the initial centroids. 

\textit{\textbf{2)}} Assign each point $\bm{x}_{i}$ ($i=1,2,\ldots,n$) to the cluster following the partition principle: if $d\left( {\bm{x}_{i},\bm{c}_{j}} \right) < d\left( {\bm{x}_{i},\bm{c}_{k}} \right),j,k = 1,2,\ldots,K,j \neq k$, $\bm{x}_{i}$ is divided into $\bm{G}_{j}$, where $d\left( {\bm{x}_{i},\bm{c}_{k}} \right)$ denotes the Euclidean distance between $\bm{x}_{i}$ and $\bm{c}_{k}$. After all the points are divided into corresponding clusters, calculate the centroids of $K$ clusters. 

\textit{\textbf{3)}} Repeat the above partition operation and centroid calculation until the centroids no longer move or a certain termination condition is met. 

K-means can be regarded as an optimization problem, and its objective function is a square error function:
{\setlength\abovedisplayskip{5pt}
\setlength\belowdisplayskip{5pt}
\begin{equation}
\label{eq_1}
f\left( {\bm{D},\bm{C}} \right) = ~{\sum\limits_{i = 1}^{n}{\min\left\{ \left\| {\bm{x}_{i} - \bm{c}_{k}} \right\|^{2} \middle| {k = 1,2,\ldots,K} \right\}}}
\end{equation}}

We generally follow the sum of squares of error (SSE)\upcite{66} method in selecting cluster numbers. The larger the number of clusters, the smaller the SSE. In calculating the SSE between the points under each cluster number and the corresponding cluster centroid, there will be a section with a high rate of decline. As the number of clusters increases later, the rate of decline will become stable. Generally, we choose the end value of the steepest section as the actual number of clusters.

However, K-means can easily become trapped around local optimal\upcite{67}, which contributes to poor performance. Due to its global optimization capability, the differential evolution algorithm (DE)\upcite{68} is an effective way to overcome this defect. 

DE is simple in structure and easy to implement, and it has high optimization efficiency, good robustness, and simple parameter setting. It is an evolutionary algorithm for solving optimization problems. The brief process of DE is developed into four steps: initialization, mutation, crossover, and selection.

\textit{\textbf{1) Initialization: }} 

Randomly generate $Np$ points as the initial population $\bm{P}^{G=1} = \left\{\bm{x}_{1}^{G=1},\bm{x}_{2}^{G=1},\ldots,\bm{x}_{Np}^{G=1} \right\}$, and each point in $\bm{P}^{G=1}$ represents a $D$-dimensional target vector.

\textit{\textbf{2) Mutation:}}

Randomly select three mutually distinct target vectors from the population $\bm{P}^{G}$ to form the mutation vector $\bm{v}$ as follows:
{\setlength\abovedisplayskip{5pt}
\setlength\belowdisplayskip{5pt}
\begin{equation}
\label{eq_2}
\bm{v}_{i}^{G} = \bm{x}_{r1}^{G} + F \cdot \left( {\bm{x}_{r2}^{G} - \bm{x}_{r3}^{G}} \right)
\end{equation}}where $G$ denotes the current generation, $F \in \lbrack 0,1\rbrack$ is the mutation factor, $i$ is an integer between $1$ and $Np$, and random integers $r1, r2, r3 \in \lbrack 1,Np\rbrack$ are distinct.

\textit{\textbf{3) Crossover: }}

This stage aims to develop the trial vector $\bm{u}$ either from the elements of the target vector or the elements of the mutation vector.
{\setlength\abovedisplayskip{5pt}
\setlength\belowdisplayskip{5pt}
\begin{equation}
\label{eq_3}
{u}_{i,j}^{G} = \left\{ \begin{matrix}
{v_{i,j}^{G}~~~~~if~{rand}_{i,j} \leq CR~or~j = j_{rand}} \\
{x_{i,j}^{G}~~~~~~~~~~~~~~~~~~~~~~~~~~~~~otherwise} \\
\end{matrix} \right.
\end{equation}}where ${rand}_{i,j} \in \lbrack 0,1\rbrack$ is a random value ($i \in \lbrack 0,Np\rbrack$, $j = 1,2,\ldots,D$), $CR \in \lbrack 0,1\rbrack$ is the crossover probability and $j_{rand} \in \lbrack 1,2,\ldots,D\rbrack$ denotes a random index.

\textit{\textbf{4) Selection: }}

This stage determines which vector can survive to the next generation $\bm{x}_{i}^{G + 1}$ by comparing the quality $\bm{u}_{i}^{G}$ with $\bm{x}_{i}^{G}$. The vector quality is determined by the objective function $f$. For a minimization problem, this step can be expressed as the following:
{\setlength\abovedisplayskip{5pt}
\setlength\belowdisplayskip{5pt}
\begin{equation}
\label{eq_4}
\bm{x}_{i}^{G + 1} = \left\{ \begin{matrix}
{\bm{u}_{i}^{G}~~~~~~~~~if~f\left( \bm{u}_{i}^{G} \right) \leq f\left( \bm{x}_{i}^{G} \right)} \\
{\bm{x}_{i}^{G}~~~~~~~~~~~~~~~~~~~~~~otherwise} \\
\end{matrix} \right.
\end{equation}}

Based on the above four stages, DE guides the direction of optimization search by the mutual cooperation and competition among the vectors in $\bm{P}$. In our work, introducing of DE can alleviate the dilemma of K-means falling into local optimal in the clustering process.

\subsection{Proposed approach}

\subsubsection{Motivation}
\noindent
Clustering data points based on distance or similarity is a powerful tool for studying many natural and social problems, such as social role identification in a specific domain. In Section 5.1, we introduce the classic K-means clustering algorithm, which is a typical unsupervised learning algorithm. Clustering algorithms put samples with similar properties into a group, and each group is also called a cluster. In the process of role identification of cyberbullying on social edge computing devices, the clustering algorithm is used to gather individuals with strong similarities into the same cluster, and the similarities and differences between these clusters can be analyzed and compared to find the roles represented by individuals. K-means is the best-known partition clustering algorithm, and its simplicity and efficiency make it the most widely used of all clustering algorithms. However, K-means can easily become trapped around local optimal\upcite{69}, contributing to poor performance. Due to its global optimization capability, DE is an effective way to overcome this defect. Therefore, we combine DE and K-means to form a clustering optimization problem, which can not only divide the data but also avoid the problem of local optimization during the clustering process.

\begin{algorithm}[h]
\renewcommand{\algorithmicrequire}{\textbf{Input:}}
\renewcommand{\algorithmicensure}{\textbf{Output:}}
\caption{\textbf{The framework of DEK}}
\label{alg1}
\begin{algorithmic}[1] 
    \STATE \textbf{Initialization phase:} \\
    $\bm{D} = \left\{ \bm{d}_{1},\bm{d}_{2},\ldots,\bm{d}_{n} \right\}$, the original clustering dataset;  
    $K$, the clustering number; 
    $MaxGs$, the maximum number of iterations;
    $Np$, the population size;
    $F$, the scaling factor;
    $Cr$, the crossover probability. 
    \STATE  $G = 1$;
    \STATE  Data continuous operation;
    \STATE  Generate initial population $\bm{P}^{G=1} = \left\{ \bm{x}_{1}^{G=1},\bm{x}_{2}^{G=1},\ldots,\bm{x}_{Np}^{G=1} \right\}$ randomly;
    \REPEAT
        \FOR { $\bm{x}_{i}^{G}$ in $\bm{P}^{G}$}
		\STATE Find the optimal individual $\bm{x}_{best}^{G}$;
		\STATE Use mutation operators to generate mutation vectors $\bm{v}_{i}^{G}$;
  		\STATE Perform the crossover operation on the target vector $\bm{x}_{i}^{G}$ and the mutation vector $\bm{v}_{i}^{G}$ using a crossover strategy to produce a trial vector $\bm{u}_{i}^{G}$;
            \STATE Perform the dimensionality reduction on the target vector $\bm{x}_{i}^{G}$ and the trial vector $\bm{u}_{i}^{G}$;
      	\STATE Evaluate the objective function values of the target vector $\bm{x}_{i}^{G}$ and the trial vector $\bm{u}_{i}^{G}$ by the objective function;
            \STATE Compare  the objective function values of the target vector $\bm{x}_{i}^{G}$ and the trial vector $\bm{u}_{i}^{G}$, and the better vector is stored in the next generation population;
            \STATE Update the optimal vector;   
        \ENDFOR 
    \UNTIL $Gs \leq MaxGs$
    \STATE   $Gs = Gs + 1$;
    \ENSURE  Different clusters and their vectors.
\end{algorithmic}  
\end{algorithm}

Many real-world datasets contain more than one type of variable. In our work, there are two types of variables, i.e., continuous and categorical variables, after modeling role features of the edge crowd in Section 4. Among these variables, categorical variables refer to sentiment polarity, emotional type, gender, and authentication. Continuous variables have a magnitude or order relationship, such as sentiment scores. For a clustering optimization problem with continuous variables and categorical variables, the existence of categorical variables will produce a large number of feasible domains, which will interfere with the solution of the problem. The continuous variable can take any value within a certain interval, while the categorical variable takes its value from a finite set of selections. For example, suppose a problem contains five categorical variables, each containing ten selections. In that case, the whole problem will become a complex optimization problem with an optimization search in 100,000 feasible domains. If categorical variables are rounded and treated as continuous variables, the presence of numeric values defaults to different selections with numeric order relationships. Because of its disorder, it will interfere with the problem’s solution, resulting in a significant deviation.

Therefore, based on the combination of K-means and DE, an improved design for the optimization process in the mixed categorical-continuous space should be carried out. We intend to change how variables are handled and introduce a distance metric different from Euclidean distance measurement. In detail, we convert the original K-means to optimizing the matrix composed of centroids, design a new optimization mechanism, and improve the objective function using the introduced distance metric. By doing this, the clustering algorithm can avoid falling into local optimal and achieve the clustering partition with higher accuracy in the mixed space. The specific implementation process of DEK is explained in detail in the following, and Algorithm 1 gives its pseudocode.

\subsubsection{Clustering optimization objective and data continuity operation}
\noindent
In this stage, we perform digital operations on the features extracted in Section 4 to make them recognized by the computer. Then, we construct a clustering problem in the mixed categorical-continuous space to determine the clustering optimization objective and carry out continuity operations on the variables. Specifically, by introducing DE into the original K-means clustering algorithm, the original clustering problem is transformed into the optimization problem of optimizing the cluster centroid. Details are as follows:

The dataset is $\bm{D} = \left\{ \bm{d}_{1},\bm{d}_{2},\ldots,\bm{d}_{n} \right\}$ in the mixed categorical-continuous space, where each data $\bm{d}_{i}(i = 1,2,\ldots,n)$ is a $m$ dimensional vector and the number of clusters is $K$. Suppose $\bm{x}_{j}(j = 1,2,\ldots,K)$ is all the centroids to be optimized, thus the decision variable is $\bm{X} = \left( \bm{x}_{1},\bm{x}_{2},\ldots,\bm{x}_{k} \right)$. The objective function to be optimized can be expressed as:
{\setlength\abovedisplayskip{5pt}
\setlength\belowdisplayskip{5pt}
\setcounter{equation}{4}
\begin{equation}
\label{eq_5}
f\left( {\bm{D},\bm{X}} \right) = {\sum\limits_{i = 1}^{n}{\min\left\{ {\varphi\left( \bm{d}_{i},\bm{x}_{j} \right)} \middle| {j = 1,2,\ldots,K} \right\}}}
\end{equation}}where $\varphi$ is a distance function. Eq. (5) aims to minimize the sum of the distances between the clustering data and the centroid of the cluster to which they belong. Eq. (6) shows that the decision variable is a $K \times m$ size centroid matrix.
{\setlength\abovedisplayskip{5pt}
\setlength\belowdisplayskip{5pt}
\begin{equation}
\label{eq_6}
\bm{X} = \begin{bmatrix}\bm{x}_{1}\\\bm{x}_{2}\\\vdots\\\bm{x}_{K}\end{bmatrix}
= \begin{bmatrix}x_{1,1}&x_{1,2}&\cdots&x_{1,m}\\
x_{2,1}&x_{2,2}&\cdots&x_{2,m}\\
\vdots&\vdots&\ddots&\vdots\\
x_{K,1}&x_{K,2}&\cdots&x_{K,m}\end{bmatrix}
\end{equation}}

Since DE is an optimization algorithm designed for continuous variables, the original algorithm must be improved in the mixed space when mixed variables appear. Assuming that $D_{con}$ and $D_{cat}$ represent the dimension of continuous and categorical variables, the number of options corresponding to each categorical variable is $N_{1},N_{2},\ldots,N_{D_{cat}}$. We transform categorical variables into continuous variables whose value range is $\lbrack0,1\rbrack$. We define these transformed continuous variables as the selection probability vector to its responding categorical variable. The length of a categorical variable’s corresponding selection probability vector equals the number of possible selections for this categorical variable. After the continuity transformation, the problem is transformed from the original $D_{con} + D_{cat} = m$ dimensions to $D_{con} + {\textstyle\sum_{l = 1}^{D_{cat}}}N_{l}$ dimensions, as shown in Eq. (7).

\begin{equation}
\label{eq_7}
\bm{X}  = \begin{bmatrix}\bm{x}_{1}\\\bm{x}_{2}\\\vdots\\\bm{x}_{K}\end{bmatrix} = \begin{bmatrix}x_{1,1}&\cdots&x_{1,D_{con}}&x_{1,D_{con}+1}&\cdots&x_{1,D_{con} + {\textstyle\sum_{l = 1}^{D_{cat}}}N_{l}}\\
x_{2,1}&\cdots&x_{2,D_{con}}&x_{2,D_{con}+1}&\cdots&x_{2,D_{con} + {\textstyle\sum_{l = 1}^{D_{cat}}}N_{l}}\\
\vdots&\ddots&\vdots&\vdots&\ddots&\vdots\\
x_{K,1}&\cdots&x_{K,D_{con}}&x_{K,D_{con}+1}&\cdots&x_{K,D_{con} + {\textstyle\sum_{l = 1}^{D_{cat}}}N_{l}}\end{bmatrix}
\end{equation}

\vspace{1ex}

\begin{equation}
\label{eq_8}
\bm{X}_{new} = \begin{bmatrix}x_{1,1}&\cdots&x_{1,D_{con} + {\textstyle\sum_{l = 1}^{D_{cat}}}N_{l}}&x_{2,1}&\cdots&x_{2,D_{con} + {\textstyle\sum_{l = 1}^{D_{cat}}}N_{l}}&\cdots&x_{K,1}&\cdots&x_{K,D_{con} + {\textstyle\sum_{l = 1}^{D_{cat}}}N_{l}}\end{bmatrix}
\end{equation}
\vspace{1ex}
\begin{equation}
\label{eq_9}
\bm{X}^{'} 
= \begin{bmatrix}\bm{x}_{1}^{'}\\\bm{x}_{2}^{'}\\\vdots\\\bm{x}_{K}^{'}\end{bmatrix}
= \begin{bmatrix}x_{1,1}&\cdots&x_{1,D_{con}}&\cdots&  p(\begin{bmatrix}{x_{1,D_{con} + {\textstyle\sum_{l = 1}^{t-1}}N_{l}+1}}&\cdots&{x_{1,D_{con} + {\textstyle\sum_{l = 1}^{t}}N_{l}}}\end{bmatrix})  &\cdots\\
                 x_{2,1}&\cdots&x_{2,D_{con}}&\cdots&  p(\begin{bmatrix}{x_{2,D_{con} + {\textstyle\sum_{l = 1}^{t-1}}N_{l}+1}}&\cdots&{x_{2,D_{con} + {\textstyle\sum_{l = 1}^{t}}N_{l}}}\end{bmatrix})   &\cdots\\
                 \vdots&\ddots&\vdots&\vdots&\ddots&\vdots\\
                x_{K,1}&\cdots&x_{K,D_{con}}&\cdots&   p(\begin{bmatrix}{x_{K,D_{con} + {\textstyle\sum_{l = 1}^{t-1}}N_{l}+1}}&\cdots&{x_{K,D_{con} + {\textstyle\sum_{l = 1}^{t}}N_{l}}}\end{bmatrix})   &\cdots\end{bmatrix}
\end{equation}

\subsubsection{Evolution and data discretization operation}
\noindent
After determining the clustering optimization objective and data continuity operation, we carry out evolution and data discretization operation on the K-means algorithm assisted by DE in the mixed categorical-continuous space to obtain the optimal classification.

To simplify the calculation, we transform the matrix in Eq. (7) into a vector form, that is, the centroid matrix with the original size of $K \times m$ changes to a vector with $1 \times \left( K \times \left( D_{con} + {\textstyle\sum_{l = 1}^{D_{cat}}}N_{l} \right) \right.$ in Eq. (8). Hence, the original problem is a whole continuous problem whose decision variable is $\bm{X}_{new}$. We subsequently perform the initialization, mutation, and crossover operations according to Eqs. (2), (3), and (4). Specifically, the vector initialization stage generates multiple uniform vectors and randomly selects them as the initial population. Next, two target vectors are randomly selected in the initial population, and the proportion difference between the two target vectors is added to the third target vector of the current population to generate the mutation vector. Then, it enters the crossover operation stage, in which the trial vector is introduced to ensure it can inherit at least one component from the target vector and the mutation vector. Finally, a greedy selection scheme is carried out to measure the degree of achieving the best performance. This stage transforms the trial and target vectors into the original mixed variable states. By comparing the quality of the trial vector obtained from the crossover operation phase with its corresponding target vector quality, we determine who can survive to the next generation. The quality of the solution is calculated using the objective function designed for the mixed variable clustering optimization problem for the current cyberbullying scenario. If the trial vector has a poor or equal fitness function value, it will replace the corresponding target vector for the next generation. Otherwise, the old target vector will remain in the next generation.

In the selection stage, three issues will be focused. In the first issue, $\bm{X}_{new}$ should be transformed into the original state for evaluation by the objective function. To one categorical variable, we take the index of the largest value in its corresponding selection probability vector as its categorical value. The transformed expression form is shown in Eq. (9), where $p$ is a function that returns the index of the position of the largest value in the vector $\bm{x}$ and $t$ is an integer selected from the set $\left\{1,2,\cdots, D_{cat}\right\}$.

The second issue in the selection stage is to introduce the distance metric. Due to the existence of mixed categorical-continuous variables, we discard the original Euclidean distance metric for continuous variables and introduce Gower distance, which can handle mixed variables\upcite{70}. Take two points $\bm{p}$ and $\bm{t}$ in the mixed space as an example. The Gower distance between two points is the average of all dimensional comparisons:
{\setlength\abovedisplayskip{5pt}
\setlength\belowdisplayskip{5pt}
\setcounter{equation}{9}
\begin{equation}
\label{eq_10}
d^{Gow}\left( {\bm{p},\bm{t}} \right) = \frac{\sum_{k = 1}^{m}s_{p_{k}t_{k}}}{m}
\end{equation}}where $m$ is the dimension. If the coordinates are categorical variables, the values of $s_{p_{k}t_{k}}$ are as follows: if $\bm{p}$ and $\bm{t}$ have the same values in the $k$-th dimension, then $s_{p_{k}t_{k}} = 0$, otherwise $s_{p_{k}t_{k}} = 1$. If it is a continuous variable coordinate, $s_{p_{k}t_{k}} = {\left| {p_{k} - t_{k}} \right|}/{R_{k}}$, $R_{k}$ is the range of $k$-th input. 

The third issue in the selection stage is the objective function design. As shown in Eq. (11), apart from the Gower distance between the clustering data and the centroid of the cluster to which they belong, we also consider the Gower distance between the centroids as one of the evaluation criteria in the objective function.
{\setlength\abovedisplayskip{5pt}
\setlength\belowdisplayskip{5pt}
\begin{equation}
\label{eq_11}
\varphi\left( {\bm{d}_{i},\bm{x}_{j}} \right) = \frac{d^{Gow}\left( {\bm{d}_{i},\bm{x}_{j}} \right)}{\ln\left\lbrack {d^{Gow}\left( {\bm{x}_{j},\bm{x}_{j}^-} \right)} \right\rbrack}
\end{equation}}where $\bm{x}_{j}^-$ denotes current centroids except for $\bm{x}_{j}$. Finally, calculate the objective function value using Eqs. (5) and (11), generate the next generation, and cycle DEK until the termination criterion is reached.

\begin{table}[!htb]\small 
\centering
\caption{\centering{{Parameter settings for DEK.}}}
\label{table1}
\vspace{2.5mm}
\begin{tabular}{p{20em}<{\centering}p{5em}<{\centering}}
\toprule
 \textbf{Parameter} & \textbf{Value}   \\ 
 \midrule
$Np$, the population size &	60	 \\
$F$, the scaling factor&	0.7	\\
$Cr$, the crossover probability&	0.8	\\
$MaxGs$, the maximum number of iterations &	1500	\\
\bottomrule
\end{tabular}
\end{table}

\begin{table}[!htb]\small 
\centering
\caption{\centering{{Descriptive statistics of the general datasets.}}}
\label{table2}
\vspace{2.5mm}
\begin{tabular}{p{9em}<{\centering}p{3em}<{\centering}p{3em}<{\centering}p{3em}<{\centering}p{11.5em}<{\centering}}
\toprule
 \textbf{Dataset} & \textbf{\# }  & $\bm{D_{con}}$  & $\bm{D_{cat}}$  & \textbf{\# choices in $\bm{D_{cat}}$}  \\ 
 \midrule
Australian	&	690	&	6	&	8	&	[2 3 14 9 2 2 2 3]		\\
Diagnosis	&	120	&	1	&	5	&	[2 2 2 2 2]		\\
Heart	&	270	&	8	&	5	&	[2 4 3 2 3]		\\
German	&	1000	&	7	&	13	&	[4 5 10 5 5 4 3 4 3 3 4 2 2]	\\
SouthGermanCredit	&	1000	&	7	&	13	&	[4 5 11 5 5 4 3 4 3 3 4 2 2]	\\
\bottomrule
\end{tabular}
\end{table}

\subsection{Experimental setup and comparison}

\subsubsection{Experimental settings}

\noindent
To assess the ability of the proposed DEK, we conducted several experiments on ten real-world datasets obtained from Weibo and five public datasets. Experiments in our work are carried out using an Intel(R) Core(TM) i7-1065G7 CPU @ 1.30GHz 1.50 GHz processor and 16 GB memory. The algorithm was programmed in MATLAB R2021b and PyCharm Community Edition 2021.3.2.

First, several well-known metrics, i.e. Davies Bouldin value (DBI), contour coefficient (SC), and Dunn index (DVI)\upcite{71}, are introduced to evaluate the performance of DEK. These three evaluation indicators do not rely on external information, such as the real label of the training sample, but only on the clustering results and the attributes of the sample itself to evaluate the clustering. Since the number of clusters in our work is unknown, using these three evaluation indicators to evaluate the effect of clustering is suitable. The smaller the DBI is, the better the clustering effect is, and the larger SC and DVI are, the better the clustering effect is. 

To explore the feasibility of the proposed DEK, we compare the DEK with seven common clustering algorithms or their variants in role identification. These compared algorithms are K-means\upcite{72}, Fuzzy C-means (FCM)\upcite{73}, K-means++\upcite{65}, K-Multiple-Means\upcite{74}, Kernel K-means\upcite{75}, GMM\upcite{76}, and Hierarchical Clustering\upcite{77}. The settings of the control parameters for DEK are detailed in Table 1. The process is repeated until the termination criterion $MaxGs$ is met.

\subsubsection{Comparison on public datasets}
\noindent
To verify the generality of DEK, we conducted experiments on five public mixed categorical-continuous datasets. The five public datasets are from the UCI Machine Learning Repository\upcite{78} and the KEEL-Dataset repository\upcite{79}. UCI currently maintains 653 datasets to serve the machine learning community, and KEEL provides a set of benchmarks for machine learning scholars to analyze the behavior of learning methods. In addition to the raw data, these dataset websites provide additional content, such as data descriptions, for scholars to use easily. Most importantly, UCI and KEEL contain many datasets with continuous and categorical variables and come from different areas of life to reflect real scenarios.

 \begin{sidewaystable}[!]\small   
\centering
 \caption{{\centering{Comparison of DEK, K-means (KM), FCM, K-means++ (KM++), K-Multiple-Means (KMM), Kernel K-means (KKM), GMM, and Hierarchical Clustering (HC) on the general dataset. The bold values represent the best performances.}}}
 \label{table3}
 \vspace{2.5mm}
\begin{tabular}{p{8.8em}<{\centering}p{3.8em}<{\centering}p{4.6em}<{\centering}p{4.6em}<{\centering}p{4.6em}<{\centering}p{4.6em}<{\centering}p{4.6em}<{\centering}p{4.6em}<{\centering}p{4.6em}<{\centering}p{4.6em}<{\centering}p{4.6em}<{\centering}p{4.6em}<{\centering}}
\toprule
\textbf{Dataset} & \textbf{Metrics} & \textbf{DEK} & \textbf{KM} &\textbf{FCM} & \textbf{KM++} & \textbf{KMM} & \textbf{KKM} & \textbf{GMM} & \textbf{HC} \\ 
\midrule
\multirow{3}{*}{Australian}	& DBI	&\textbf{1.4336 }		&	1.9801	&	2.1550	&	1.9973	&	1.6548	&	2.0254	&	2.0739	&	1.8834	\\
		& SC	&\textbf{0.2107 }		&	0.1880	&	0.1922	&	0.1874	&	0.0776	&	0.1608	&	0.1344	&	0.1951	\\
		& DVI	&	0.1123	&	0.1524	&	0.0874	&	0.1652	&	0.1632	&	0.1275	&	0.147	&	\textbf{0.3495} 	\\
\multirow{3}{*}{Diagnosis}	& DBI	&\textbf{0.3947}		&	1.2465	&	1.4577	&	1.2336	&	1.0809	&	1.2400	&	0.8644	&	0.8667	\\
		& SC	&\textbf{0.6796 }		&	0.4221	&	0.5191	&	0.4292	&	0.5002	&	0.4377	&	0.6272	&	0.5764	\\
		& DVI	&\textbf{ 0.4167}		&	0.0099	&	0.0096	&	0.0093	&	0.1379	&	0.1070	&	0.3398	&	0.3600	\\
\multirow{3}{*}{Heart}	& DBI	&\textbf{ 0.8973}		&	2.3420	&	2.3429	&	2.3432	&	1.7521	&	2.5081	&	1.9769	&	0.9387	\\
		& SC	&\textbf{0.2653 }		&	0.0509	&	0.0511	&	0.0507	&	0.2511	&	0.0070	&	0.2357	&	0.0920	\\
		& DVI	&\textbf{ 0.2680}		&	0.0256	&	0.0232	&	0.0256	&	0.0823	&	0.0134	&	0.1329	&	0.2258	\\
\multirow{3}{*}{German}	& DBI	&\textbf{1.4669 }		&	2.0267	&	2.1047	&	2.0341	&	2.0531	&	2.0116	&	2.0239	&	1.9939	\\
		& SC	&\textbf{ 0.3710}		&	0.1299	&	0.1158	&	0.1290	&	0.1260	&	0.1272	&	0.1338	&	0.1940	\\
		& DVI	&\textbf{ 0.2884}		&	0.1038	&	0.0875	&	0.1017	&	0.0968	&	0.0973	&	0.1461	&	0.2426	\\
\multirow{3}{*}{SouthGermanCredit}	& DBI	&\textbf{ 0.9743}		&	2.0518	&	2.1192	&	2.0259	&	2.0942	&	2.0346	&	2.0414	&	0.9754	\\
		& SC	&\textbf{0.2529 }		&	0.1201	&	0.1174	&	0.1273	&	0.1259	&	0.1249	&	0.1088	&	0.2229	\\
		& DVI	&	0.2160	&	0.0815	&	0.0692	&	0.0842	&	0.0842	&	0.0833	&	0.1064	&	\textbf{0.3694} 	\\
\bottomrule
\end{tabular}
\end{sidewaystable}

\begin{table}[!h]\small  
 \centering
\caption{\centering{{Descriptive statistics of the real-world datasets from ten scenarios.}}}
\label{table4}
\vspace{2.5mm}
\begin{tabular}{p{6em}<{\centering}p{5em}<{\centering}p{7em}<{\centering}p{3em}<{\centering}p{3em}<{\centering}p{9.5em}<{\centering}}
\toprule
 \textbf{Dataset} & \textbf{\# Users} & \textbf{\# Comments} & $\bm{D_{con}}$  & $\bm{D_{cat}}$  & \textbf{\# choices in $\bm{D_{cat}}$}  \\ 
 \midrule
Scenario 1&	272&	167179&	17&	4&	[ 3 7 2 2 ] \\
Scenario 2&	354&	210682&	17&	4&	[ 3 7 2 2 ]\\
Scenario 3&	664&	306067&	17&	4&	[ 3 7 2 2 ]\\
Scenario 4&	260&	64157&	17&	4&	[ 3 7 2 2 ]\\
Scenario 5&	266&	53549&	17&	4&	[ 3 7 2 2 ]\\
Scenario 6&	405&	218451&	17&	4&	[ 3 7 2 2 ]\\
Scenario 7&	294&	57509&	17&	4&	[ 3 7 2 2 ]\\
Scenario 8&	286&	152250&	17&	4&	[ 3 7 2 2 ]\\
Scenario 9&	224&	139586&	17&	4&	[ 3 7 2 2 ]\\
Scenario 10&	206&	103591&	17&	4&	[ 3 7 2 2 ]\\
Total&	3231&	1473021&	17 &	4 &	[ 3 7 2 2 ] \\
\bottomrule
\end{tabular}
\end{table}

\begin{sidewaystable}[!]\small  
\centering
\caption{{{Comparison of DEK, K-means (KM), FCM, K-means++ (KM++), K-Multiple-Means (KMM), Kernel K-means (KKM), GMM, and Hierarchical Clustering (HC) on the real-world datasets. The bold values represent the best performances.}}}
\label{table5}
\vspace{2.5mm}
\begin{tabular}{m{3.5em}<{\centering}m{3em}<{\centering}m{3.5em}<{\centering}m{3.5em}<{\centering}m{3.5em}<{\centering}m{3.5em}<{\centering}m{3.5em}<{\centering}m{3.5em}<{\centering}m{3.5em}<{\centering}m{3.5em}<{\centering}m{3.5em}<{\centering}m{3.5em}<{\centering}}
\toprule
\textbf{Methods} & \textbf{Metrics} & \textbf{Scenario 1} & \textbf{Scenario 2} & \textbf{Scenario 3} & \textbf{Scenario 4} & \textbf{Scenario 5} & \textbf{Scenario 6} & \textbf{Scenario 7} & \textbf{Scenario 8} & \textbf{Scenario 9} & \textbf{Scenario 10}  \\ 
\midrule
\multirow{3}{*}{DEK}  & DBI   &\textbf{1.0600}   & 1.2633     & \textbf{1.1375}     & \textbf{1.2309}       & \textbf{1.1386}            & 1.9029             & 1.4188      & 1.4913             &  \textbf{1.0967}            &    \textbf{1.2019}            \\
                      & SC    &\textbf{0.4559}   & \textbf{0.3153}  & \textbf{0.5057}  & \textbf{0.3843}   &\textbf{0.4393}       & \textbf{0.3703} & \textbf{0.4228}      &  \textbf{0.3633}            &  \textbf{0.4333}            &   \textbf{0.4285}             \\
                      & DVI   &\textbf{0.0359}   & 0.0265    & \textbf{0.0360}   & \textbf{0.0482}    & \textbf{0.0215}       &\textbf{0.0190}     &\textbf{0.0807}      &  \textbf{0.0373}            &    \textbf{0.0266}          &    \textbf{0.0556}            \\
\multirow{3}{*}{KM}   & DBI              & 1.8704             & 1.1366             & 1.4715             & 1.6554             & 1.4931             & 1.4284             & 1.6809             & 1.7633             & 1.5038             & 1.6704               \\
                      & SC               & -0.0696            & 0.2620              & 0.0166             & -0.0486            & 0.0781             & -0.0156            & 0.0394             & -0.0721            & -0.0946            & -0.0046              \\
                      & DVI              & 0.0010              & 0.0414             & 0.0050              & 0.0129             & 0.0042             & 0.0002             & 0.0167             & 0.0025             & 0.0017             & 0.0041               \\
\multirow{3}{*}{FCM}  & DBI              & 1.8870              & 1.5404             & 1.8034             & 1.6795             & 1.6102             & 1.4008             & 2.1085             & 2.1371             & 1.5600               & 2.7799               \\
                      & SC               & -0.0266            & 0.1091             & -0.0127            & -0.0910             & 0.0174             & -0.0305            & 0.0158             & -0.0413            & -0.0586            & -0.0309              \\
                      & DVI              & 0.0007             & 0.0068             & 0.0010              & 0.0114             & 0.0015             & 0.0002             & 0.0073             & 0.0018             & 0.0013             & 0.0044               \\
\multirow{3}{*}{KM++} & DBI              & 1.9820              & 1.3108             & 1.7069             & 1.7880              & 1.6184             & 1.7045             & 1.9217             & 2.0008             & 1.6017             & 1.7079               \\
                      & SC               & -0.0506            & 0.2397             & -0.0123            & -0.0314            & 0.0456             & -0.0158            & 0.0492             & -0.0407            & -0.0180             & -0.0062              \\
                      & DVI              & 0.0008             & 0.0302             & 0.0026             & 0.0113             & 0.0034             & 0.0002             & 0.0131             & 0.0018             & 0.0012             & 0.0040                \\
\multirow{3}{*}{KMM}  & DBI              & 2.1656             & 2.0860              & 2.0433             & 2.1312             & 2.1356             & 2.1411             & 1.9524             & 2.1205             & 2.0939             & 2.1349               \\
                      & SC               & -0.0790             & -0.0316            & -0.0348            & -0.0613            & -0.0470             & -0.0739            & 0.0809             & -0.0884            & -0.0649            & -0.0138              \\
                      & DVI              & 0.0007             & 0.0012             & 0.0007             & 0.0051             & 0.0005             & 0.0000            & 0.0160              & 0.0016             & 0.0004             & 0.0029               \\
\multirow{3}{*}{KKM}  & DBI              & 2.4365             & 2.3968             & 2.3030              & 2.4360              & 2.3429             & 2.2885             & 2.3428             & 2.4087             & 2.3516             & 2.3728               \\
                      & SC               & -0.0672            & -0.0512            & -0.0595            & -0.0635            & -0.0534            & -0.0725            & -0.0404            & -0.0625            & -0.0712            & -0.0627              \\
                      & DVI              & 0.0000             & 0.0008             & 0.0003             & 0.0035             & 0.0002             & 0.0000                  & 0.0015             & 0.0014             & 0.0001             & 0.0017               \\
\multirow{3}{*}{GMM}  & DBI              & 1.4714             & 1.6773             & 1.6998             & 1.5524             & 1.6177             & 1.5644             & 1.6925             & 1.7501             & 1.6048             & 1.6212               \\
                      & SC               & 0.3035             & 0.1951             & 0.2356             & 0.2137             & 0.2048             & 0.2132             & 0.2866             & 0.2171             & 0.2179             & 0.2327               \\
                      & DVI              & 0.0150              & 0.0075             & 0.0154             & 0.0181             & 0.0066             & 0.0064             & 0.0295             & 0.0243             & 0.0074             & 0.0254               \\
\multirow{3}{*}{HC}   & DBI              & 1.2528             &  \textbf{0.8838}            & 1.1546             & 1.4144             & 1.2028             &  \textbf{1.1742}            &   \textbf{1.0572}           &  \textbf{1.3655}            & 1.1914             & 1.5105               \\
                      & SC               & -0.3657            & 0.1180              & 0.2137             & -0.3512            & -0.0007            & 0.0493             & -0.0634            & -0.1113            & -0.2659            & -0.0810               \\
                      & DVI              & 0.0121             &  \textbf{0.0532}            & 0.0346             & 0.0301             & 0.0103             & 0.0111             & 0.0590              & 0.0191             & 0.0073             & 0.0338               \\
\bottomrule
\end{tabular}
\end{sidewaystable}

 \begin{table}[!]\small 
\centering
 \caption{{{Comparison of algorithm performance by metric.}}}
 \label{table6}
 \vspace{2.5mm}
\begin{tabular}{m{7em}<{\centering}m{7em}<{\centering}m{18em}<{\centering}}
\toprule
\textbf{Method} & \textbf{Metrics} & {\textbf{\# best performance according to the metrics}  } \\ 
\midrule
\multirow{3}{*}{DEK}& DBI	&{11}		\\
		          & SC	&{15}		\\
		          & DVI	&{12}		\\
\multirow{3}{*}{HC}	& DBI	&{4}		\\
		          & SC	&{0}		\\
		          & DVI	&{3}		\\
\bottomrule
\end{tabular}
\end{table}

In these public datasets, categorical variables take values from a finite set of selections, which usually identify non-numeric elements of an unordered set. For example, in the German Credit Data dataset, the variable Purpose contains ten non-numeric elements such as car (new), car (used), furniture/equipment, radio/television, etc. The number of selections in the variable Purpose is 10. It is worth mentioning that some of these datasets contain integer variables. There is a specific natural order relationship between these integer variables, and the processing of such variables can use the traditional continuous variable processing method, so we treat them as continuous variables. Descriptive data for these five public datasets is shown in Table 2, and Table 3 presents the comparative experimental results over 20 independent runs. It is clear that DEK achieves significantly better results than the other algorithms on five public mixed categorical-continuous datasets.

\subsubsection{Comparison on real-world datasets}
\noindent
After comparative experiments on public datasets, we found that the proposed DEK effectively clusters on mixed categorical-continuous variable data. Next, we used ten real-world cyberbullying datasets from Section 3 to evaluate the algorithm’s performance according to DBI, SC, and DVI. Based on data characteristics of the edge crowd in Section 4, we defined seventeen continuous variables and four categorical variables. The statistics of real-world cyberbullying datasets are summarized in Table 4. Note that the ten real-world datasets in Table 4 correspond to the ten cyberbullying scenarios in Section 3. Table 5 presents the DEK’s DBI, SC, and DVI performance over 20 independent runs. We found that DEK performs better than other algorithms in real-world mixed categorical-continuous datasets. To express the algorithm comparison more intuitively, we present Table 6 to show the number of times DEK and HC achieved the best performance by DBI, SC, and DVI. It is clear that DEK achieves significantly better results than the other algorithms in relation to most metrics. Unlike other clustering algorithms, the HC can build a hierarchical structure to display the clustering results. It can handle any data type, including categorical, numerical, and mixed data. The HC can adapt to different data characteristics and clustering requirements through different distance measurement methods and aggregation strategies. Therefore, the HC in this paper performs better than the proposed DEK on some datasets or indicators.

\section{Empirical Analysis of Roles}
\noindent
In this section, we utilize DEK to aggregate role features into clusters, and the number of clusters is determined using the SSE method. By analyzing clustering members who show similar patterns of behaviors, we obtain nine typical roles under different scenarios in edge AI applications on social edge computing devices.

\begin{figure}[htbp]
\centering
\includegraphics[width=3in]{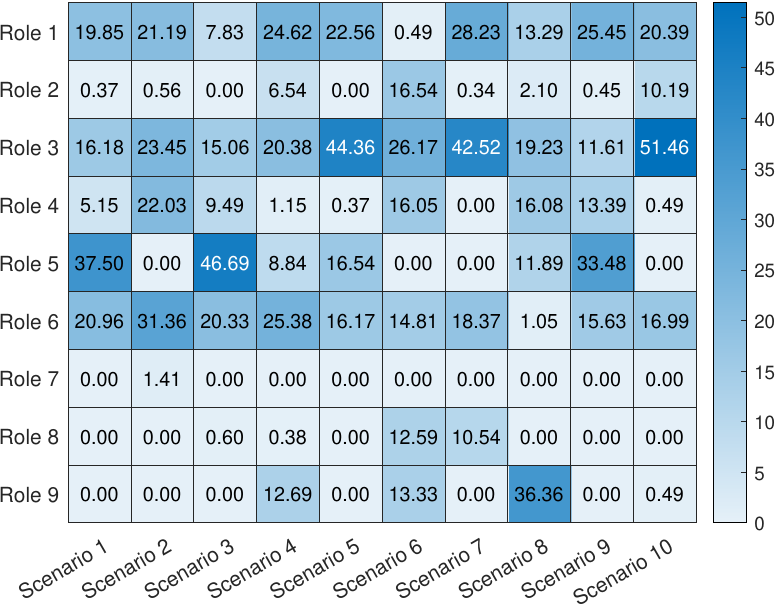}
\caption{{Distribution of occupied number of roles involved in cyberbullying.}}
\label{Fig1}
\end{figure}

\subsection{Derived roles in all scenarios}
\noindent
To answer RQ1, we first observe the edge crowd in clusters obtained by DEK and then summarize the representative behavior characteristics of nine types of roles. Fig. 1 shows the percentage of different role occupations. 

\textbf{1) Zealous Perpetrator (Role 1):} This role refers to people who tend to fill their comments with overtly insulting words and extreme emotional outbursts.

\textbf{2) Spreader of Further Escalation (Role 2):} People in this role use their influence to amplify their words, making their comments widely disseminated and discussed.

\textbf{3) Emotionally Controlled Perpetrator (Role 3):} Although people in this role use some uncivilized words, they do not unleash their grievances entirely without restraint.

\textbf{4) Encouraging Bystander (Role 4):} People in this role provide emotional support and post positive words symbolizing encouragement, which can give victims spiritual comfort.

\textbf{5) Exaggerated and Fueled Bystander (Role 5):} This role refers to people whose texts are full of cynicism or embellishment to the negative, attempting to bring the topic into a direction that is not conducive to the victim.

\textbf{6) Calm Observer Analyst (Role 6):} This role refers to people who provide dispassionate analysis of the points of view involved in the current scenario.

\textbf{7) Bystander who Meets Popular Expectations (Role 7):} This role refers to people whose commentary is dispassionate based on the direction of the situation to come up with the public’s future expectations.

\textbf{8) Perpetrator with a Purpose (Role 8):} This role refers to people whose comments are full of prominent insulting words, and their historical posts are similar to the current topic.

\textbf{9) Sympathetic Bystander (Role 9):} People in this role usually sympathize with the victims.

\subsection{Role analysis in different scenarios}

\noindent
In order to analyze the changes in the distribution and quantity of roles under different scenarios, we present changes in the distribution of people occupied in different roles over time in Fig. 2. To capture the detailed changes when cyberbullying occurs, we zoom in on the shaded parts in Fig. 2 and display the zoomed-in version in Figs. 3 and 4. Among these roles, negative roles refer to zealous perpetrators, emotionally controlled perpetrators, exaggerated and fueled bystanders, and perpetrators with a purpose. These negative roles contain unpleasant information and negative emotions. In addition to negative roles, encouraging bystanders, calm observer analysts, and sympathetic bystanders belong to positive roles that include many positive emotions or dispassionate opinions. In the rest of this subsection, we describe the changes in the distribution and quantity of roles based on clustering results and key time points in ten scenarios.

\begin{figure*} [htbp]
    \centering
  \subfloat[\label{Fig2a}]{%
       \includegraphics[width=0.32\linewidth]{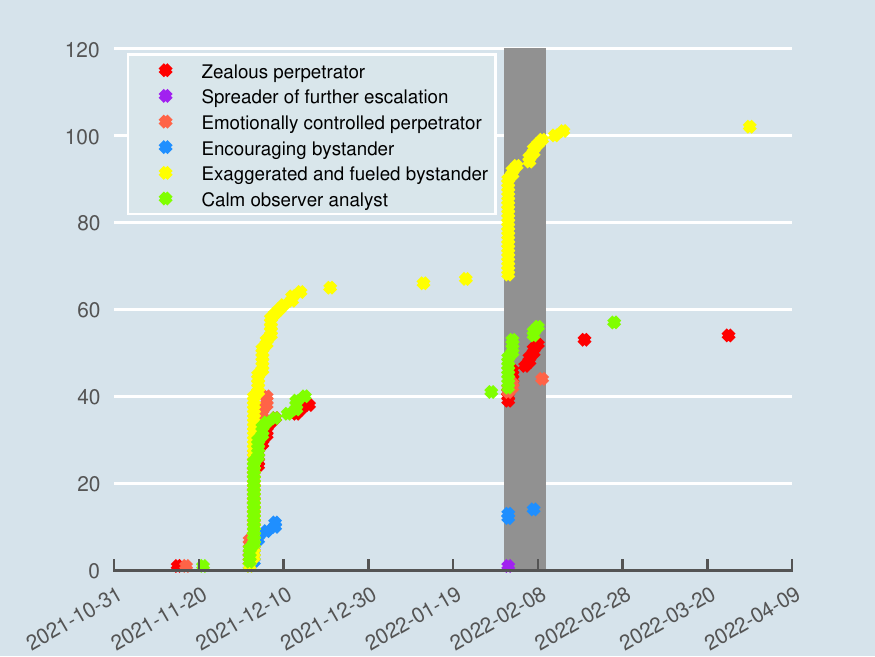}}\hspace{0.2mm}
  \subfloat[\label{Fig2b}]{%
        \includegraphics[width=0.32\linewidth]{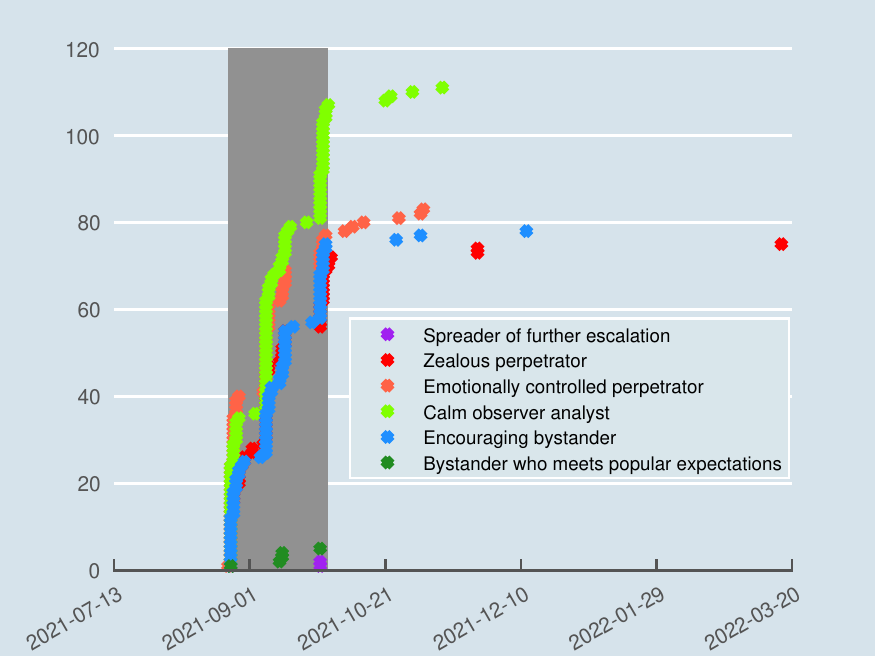}}\hspace{0.2mm}
  \subfloat[\label{Fig2c}]{%
        \includegraphics[width=0.32\linewidth]{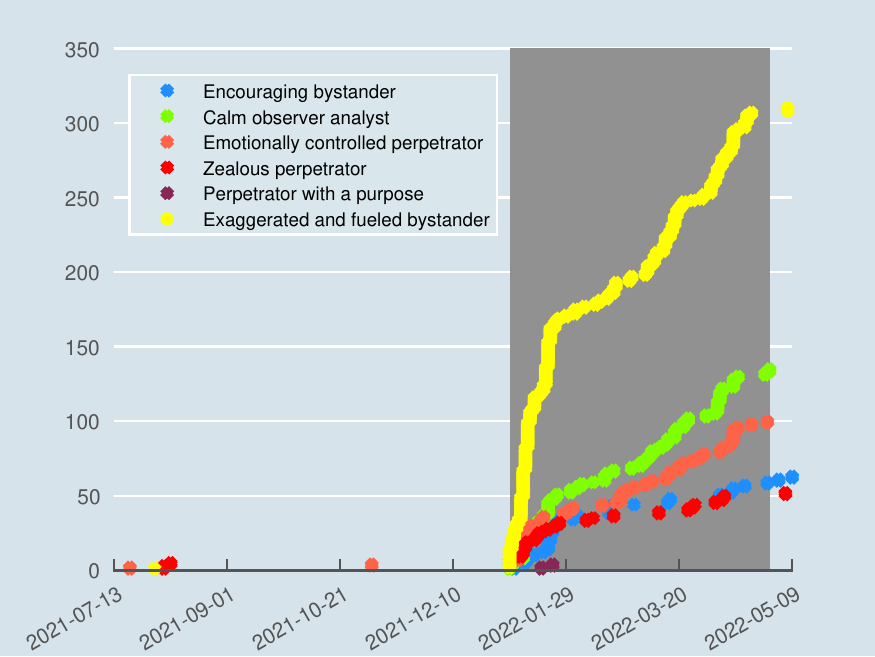}}\hspace{0.2mm}
    \\
    \vspace*{-.05in}     
  \subfloat[\label{Fig2d}]{%
        \includegraphics[width=0.32\linewidth]{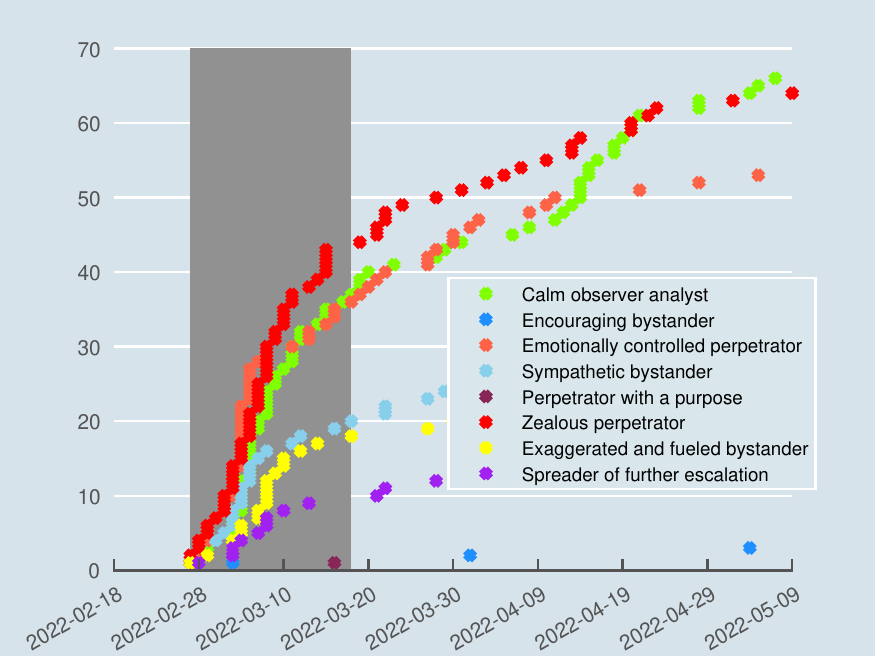}}\hspace{0.2mm}    
  \subfloat[\label{Fig2e}]{%
       \includegraphics[width=0.32\linewidth]{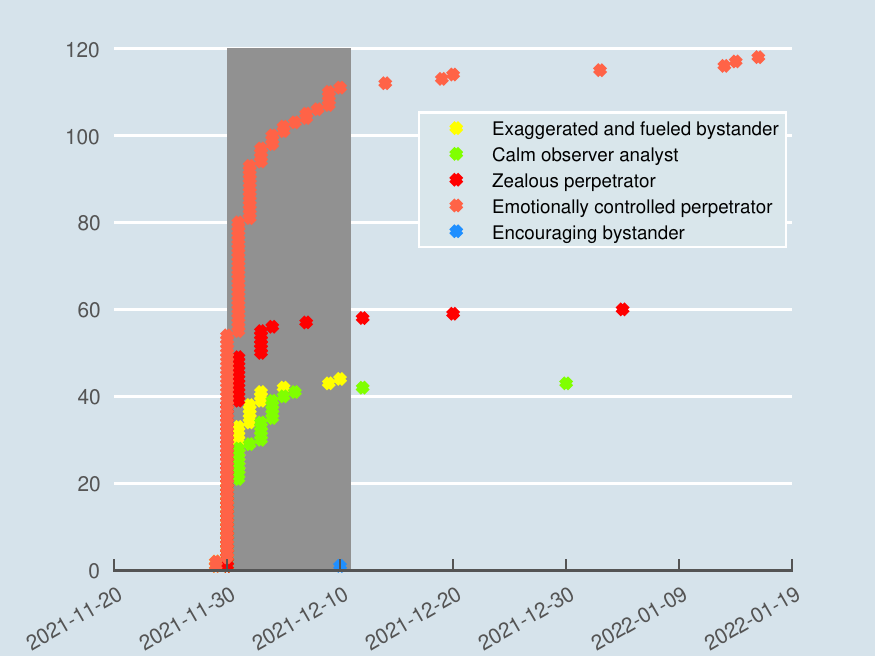}}\hspace{0.2mm}
  \subfloat[\label{Fig2f}]{%
        \includegraphics[width=0.32\linewidth]{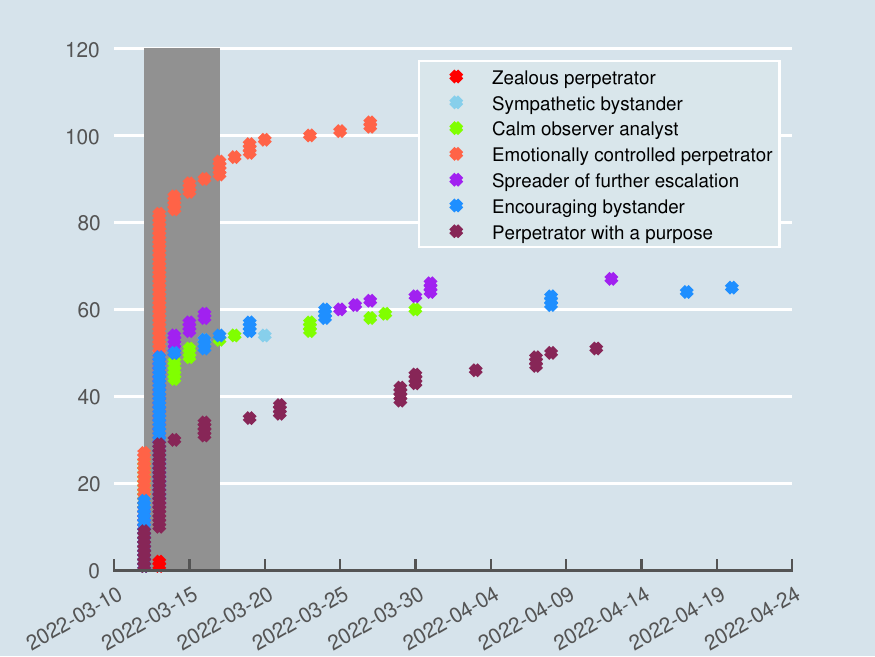}}\hspace{0.2mm} 
          \\
    \vspace*{-.05in}          
  \subfloat[\label{Fig2g}]{%
        \includegraphics[width=0.32\linewidth]{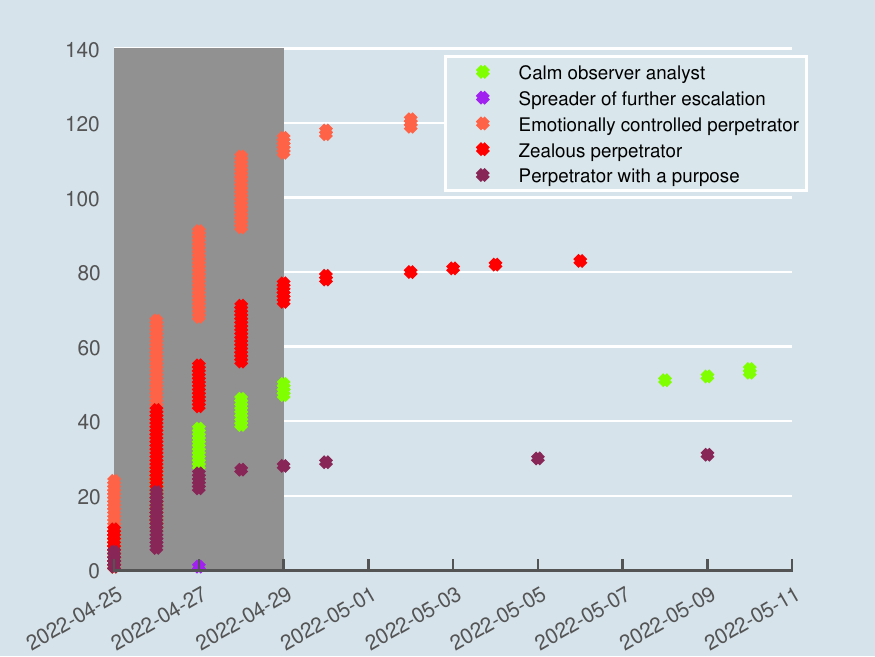}}\hspace{0.2mm}
  \subfloat[\label{Fig2h}]{%
        \includegraphics[width=0.32\linewidth]{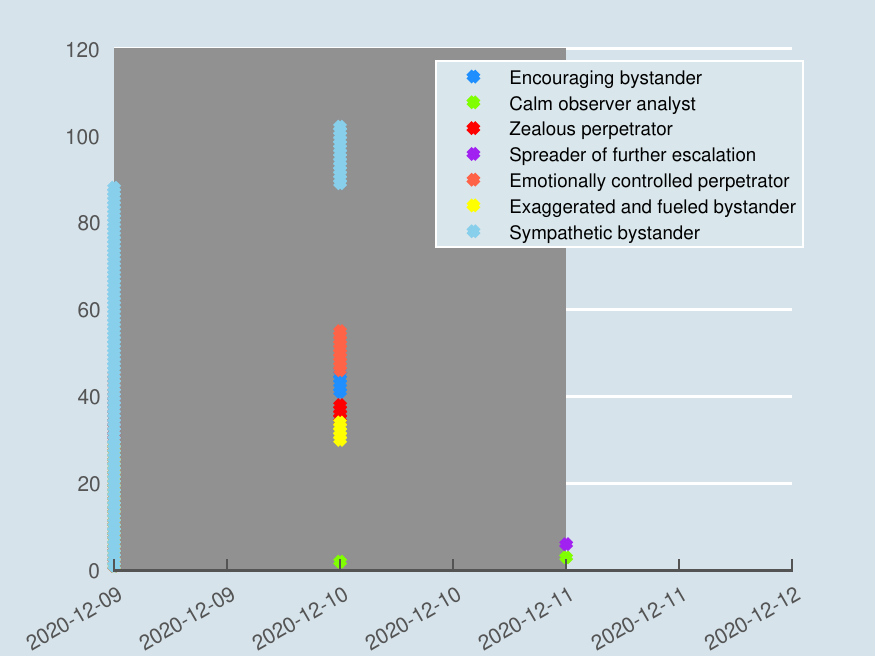}}\hspace{0.2mm}     
  \subfloat[\label{Fig2i}]{%
       \includegraphics[width=0.32\linewidth]{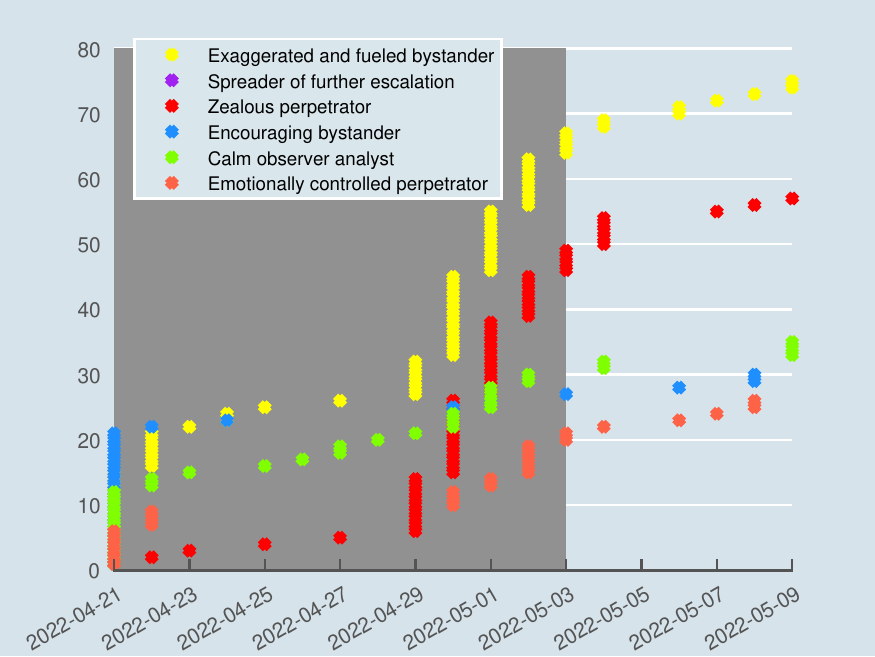}}\hspace{0.2mm}  
          \\
    \vspace*{-.05in}    
  \subfloat[\label{Fig2j}]{%
        \includegraphics[width=0.32\linewidth]{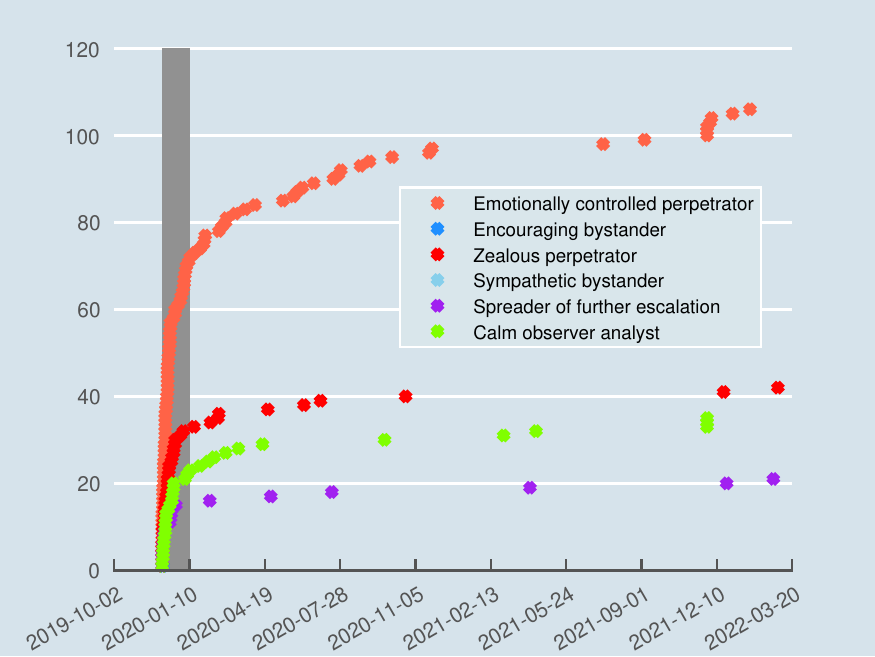}}
  \caption{\centering{Changes in the number of members occupied in roles over time in ten scenarios. (a) Scenario 1, (b) Scenario 2, (c) Scenario 3, (d) Scenario 4, (e) Scenario 5, (f) Scenario 6, (g) Scenario 7, (h) Scenario 8, (i) Scenario 9, (j) Scenario 10. The abscissa represents time while the ordinate represents the percentage of people occupied in different roles.}}
  \label{Fig2} 
\end{figure*}

\begin{figure*}[htbp]
\centering
\subfloat[]{\includegraphics[width=6.7in]{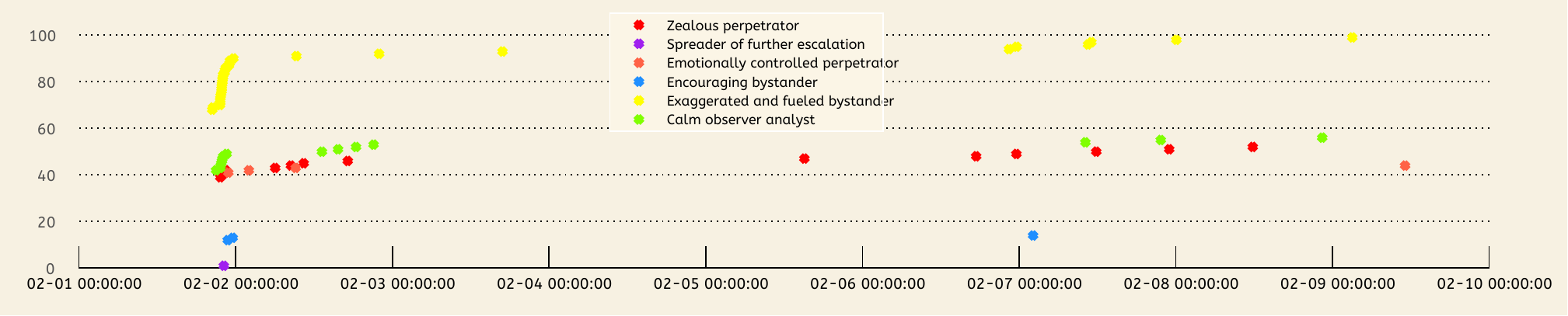}%
\label{Fig3a}}
    \vspace*{-.03in}
\hfil
\subfloat[]{\includegraphics[width=6.7in]{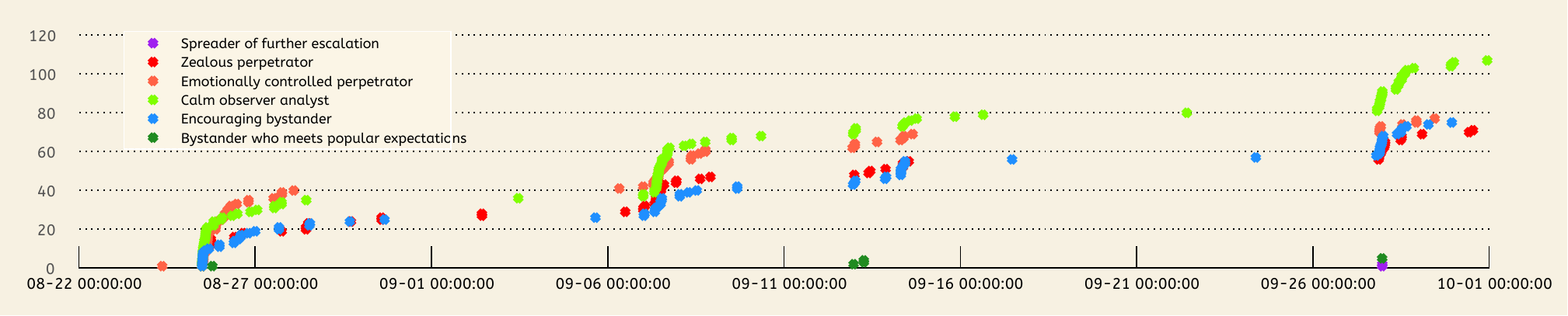}%
\label{Fig3b}}
    \vspace*{-.03in}
\hfil
\subfloat[]{\includegraphics[width=6.7in]{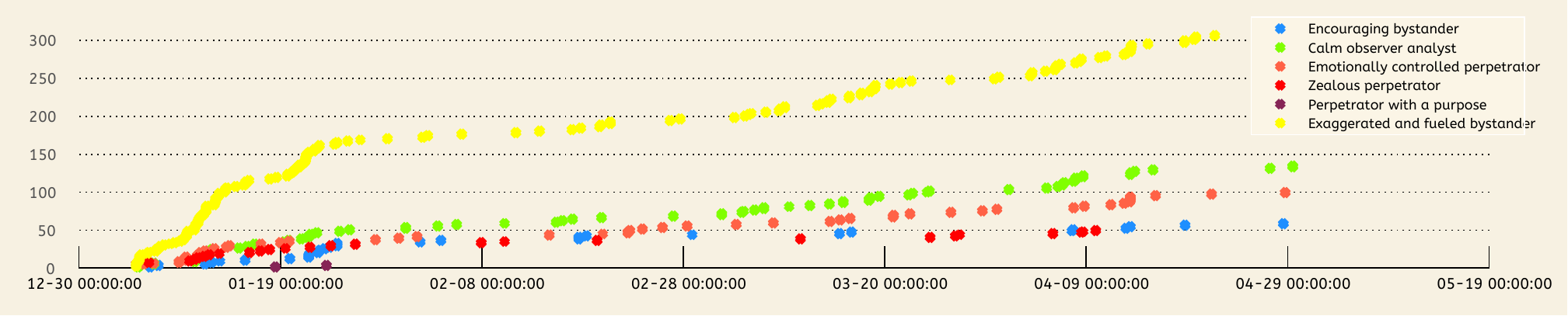}%
\label{Fig3c}}
    \vspace*{-.03in}
\hfil
\subfloat[]{\includegraphics[width=6.7in]{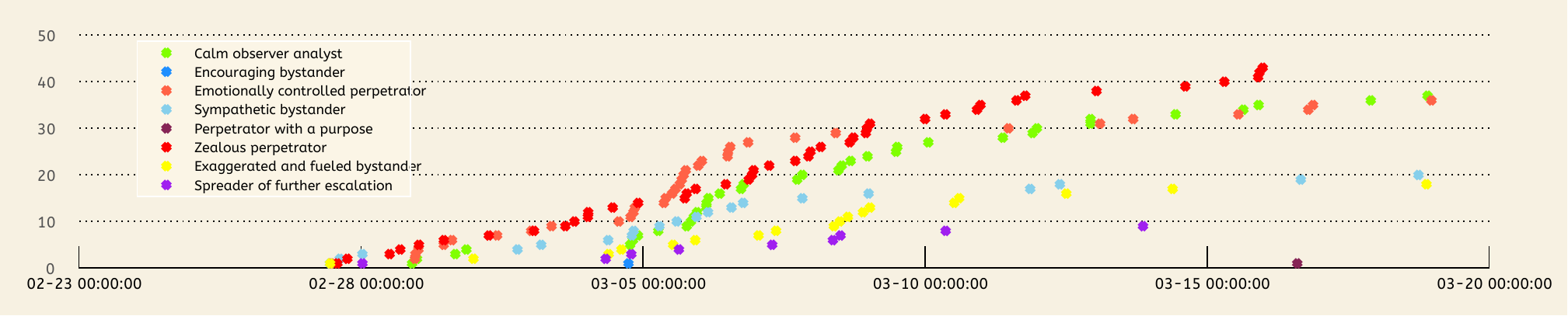}%
\label{Fig3d}}
    \vspace*{-.03in}
\hfil
\subfloat[]{\includegraphics[width=6.7in]{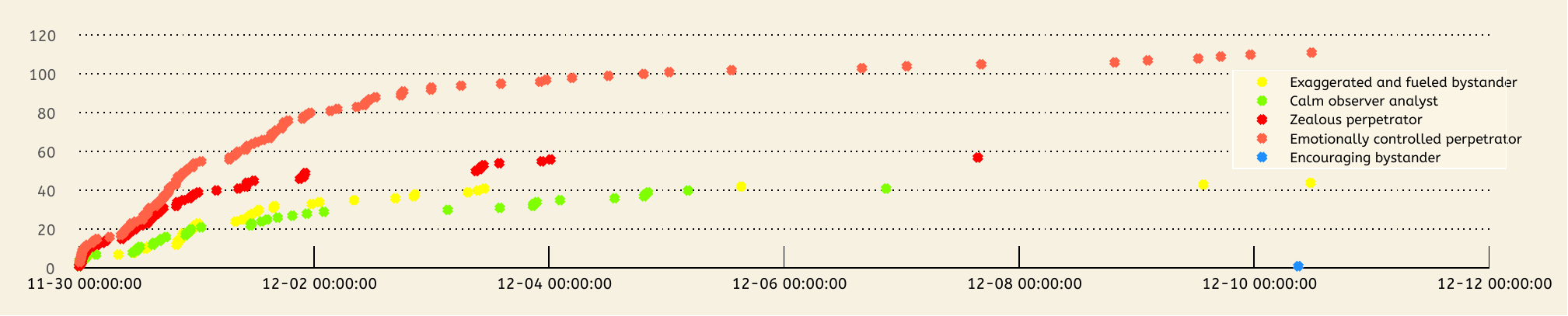}%
\label{Fig3e}}
\caption{\centering{Detail of the shaded section in Fig. \protect\ref{Fig2}\protect\subref{Fig2a} - \protect\ref{Fig2}\protect\subref{Fig2e}. (a) Scenario 1, (b) Scenario 2, (c) Scenario 3, (d) Scenario 4, (e) Scenario 5. The abscissa represents time while the ordinate represents the number of people occupied in different roles.}}
\label{Fig3}
\end{figure*}

\textbf{1) The Tie Li Case (Scenario 1)}

In Fig. 1, more than half of the members are exaggerated and fueled bystanders, calm observer analysts, and zealous perpetrators. Analyzing their historical online community behaviors, i.e., their historical Weibo profiles, shows they are more concerned about Chinese football. Some can analyze it calmly, while others are also angry because of the Chinese men’s football team’s long-term poor performance in kinds of world football games, which does not match fans’ expectations.

In Fig. 2(a), we can see that the sharp increase in the number of people represented by different roles is mainly concentrated in two phases, namely early December 2021 and early February 2022. In the first stage, the victim mainly experienced a poor record in the third round of AFC matches for the 2022 FIFA World Cup qualification and resigned from the head coach of the Chinese national football team. We focus on the second phase, shaded in Fig. 2(a), which is the cyberbullying suffered by the victim after he left the national football team.

On the evening of February 1, 2022, the Chinese national football team lost to the lowly Vietnam team. Therefore, as we can see in Fig. 3(a), the number of people represented by the different roles started to grow from the evening of February 2. Among them, in addition to the neutral and objective views expressed by calm observer analysts, more were filled with negative emotions composed of exaggerated and fueled bystanders, zealous perpetrators, and emotionally controlled perpetrators. Because of the Chinese New Year and the failure of the national football team, many fans were dissatisfied with the former coach Li and released their emotions on him. We can also find from Fig. 3(a) that these negative roles continued to appear from February 1 to February 9, reflecting netizens’ continuous attention to this incident.

\textbf{2) The Alibaba Sexual Assault Case (Scenario 2)}

In Fig. 1, most members are calm observer analysts, emotionally controlled perpetrators, encouraging bystanders, and zealous perpetrators in this scenario. In this case, a female employee of Alibaba was the first to complain about the assault by two male employees on the social media platform. Before the truth of the incident came to light, public opinion was mainly oriented to encourage and sympathize with this female employee. Given the public opinion environment at that time, the attacks on the wives of the two male employees were relatively large. However, as the wives of the two male employees clarified on Weibo and the police intervened, the case gradually came to light. As the truth continued to emerge, public opinion shifted toward the wives of the two male employees. As a result, encouraging bystanders and calm observer analysts dominated the roles as incidents unfolded.

In Fig. 2(b), we find that the sharp increase in the number of people represented by different roles is mainly concentrated from late August to late September 2021. We focus our analysis on the shaded areas in Fig. 2(b), as shown in Fig. 3(b). The increase occurred at four distinct time points: August 26, September 6, September 13, and September 27. The incidents at these four points involved the voice of the involved wife’s microblog and the police blotter, which could quickly form a significant public opinion. The cut-off point for the change in the number of members of different roles was on September 6, when the police announced the latest investigation results that Wang was detained for 15 days, and the police stopped the investigation. This move implied that although the allegations made by the female employee of Alibaba were true, they were exaggerated. Thus, after that, the number of calm observer analysts and encouraging bystanders outnumbered the number of zealous perpetrators and emotionally controlled perpetrators. In the fourth stage in Fig. 3(b), that is, around September 27, influenced by the results of the previous investigation by the police and the disseminators of the spreaders of further escalation, the number of calm observer analysts is significantly higher than that of other members with negative roles.

\begin{figure*}[htbp]
\centering
\subfloat[]{\includegraphics[width=6.7in]{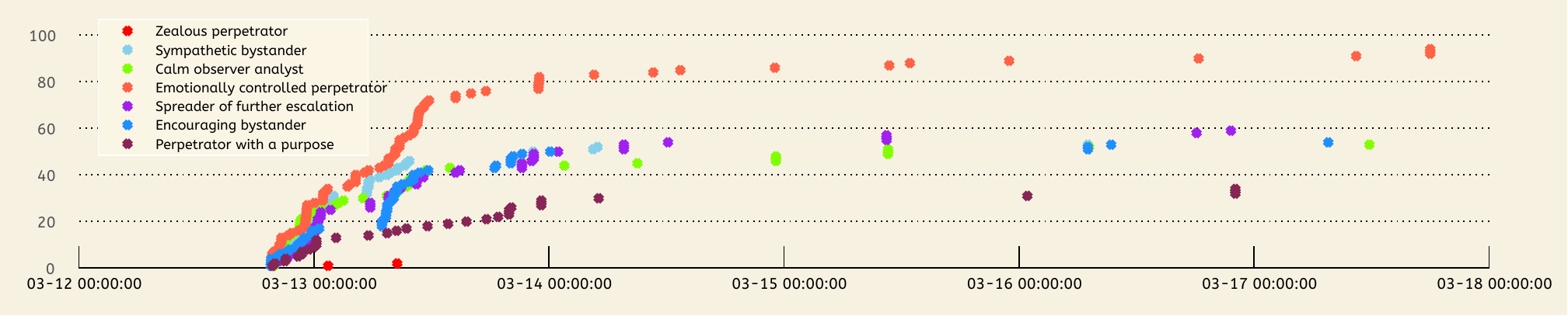}%
\label{Fig4a}}
    \vspace*{-.03in}
\hfil
\subfloat[]{\includegraphics[width=6.7in]{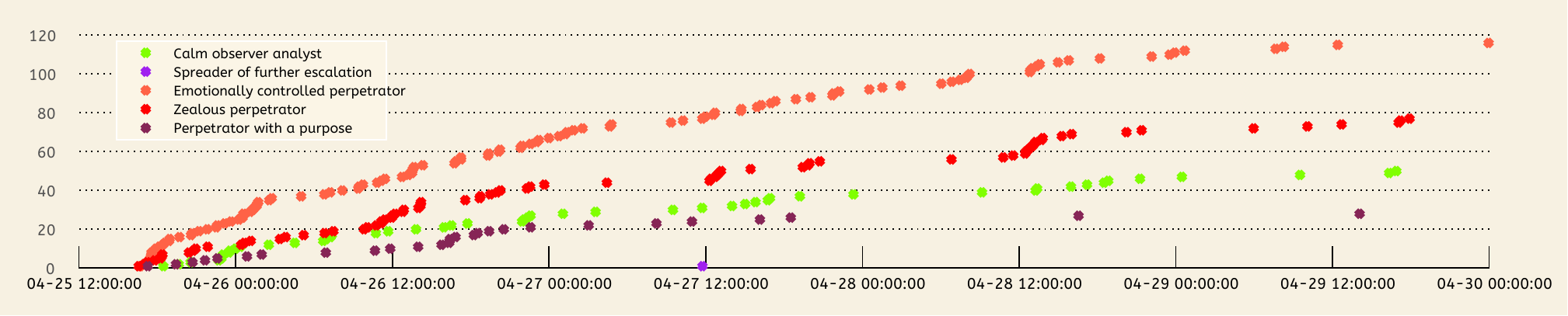}%
\label{Fig4b}}
    \vspace*{-.03in}
\hfil
\subfloat[]{\includegraphics[width=6.7in]{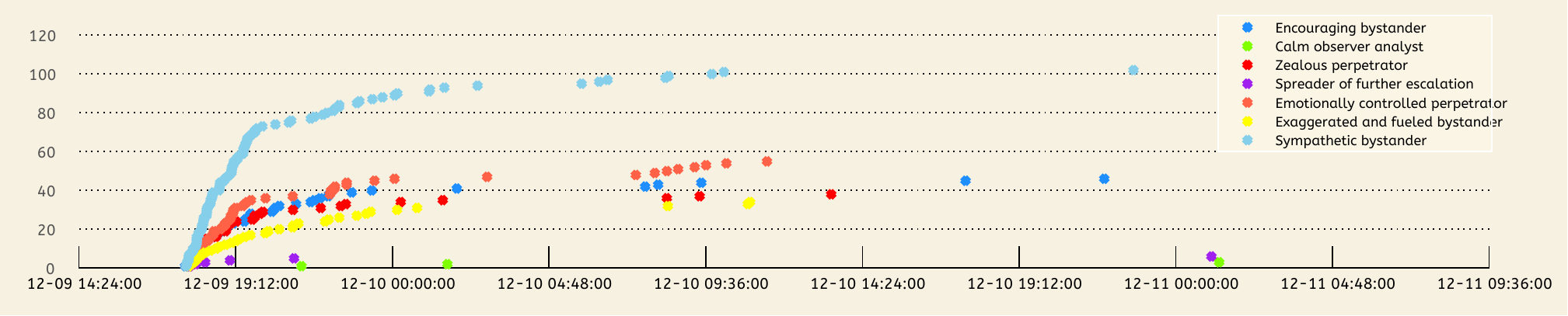}%
\label{Fig4c}}
    \vspace*{-.03in}
\hfil
\subfloat[]{\includegraphics[width=6.7in]{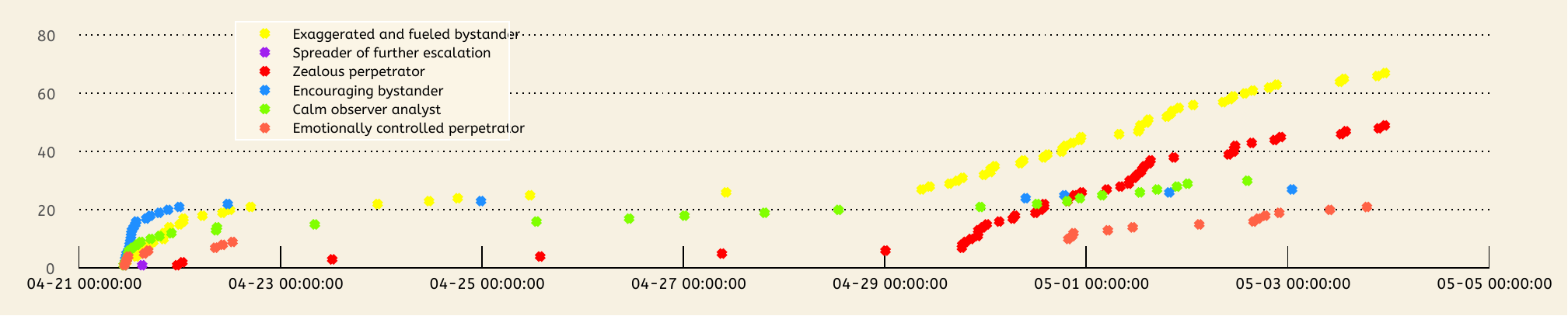}%
\label{Fig4d}}
    \vspace*{-.03in}
\hfil
\subfloat[]{\includegraphics[width=6.7in]{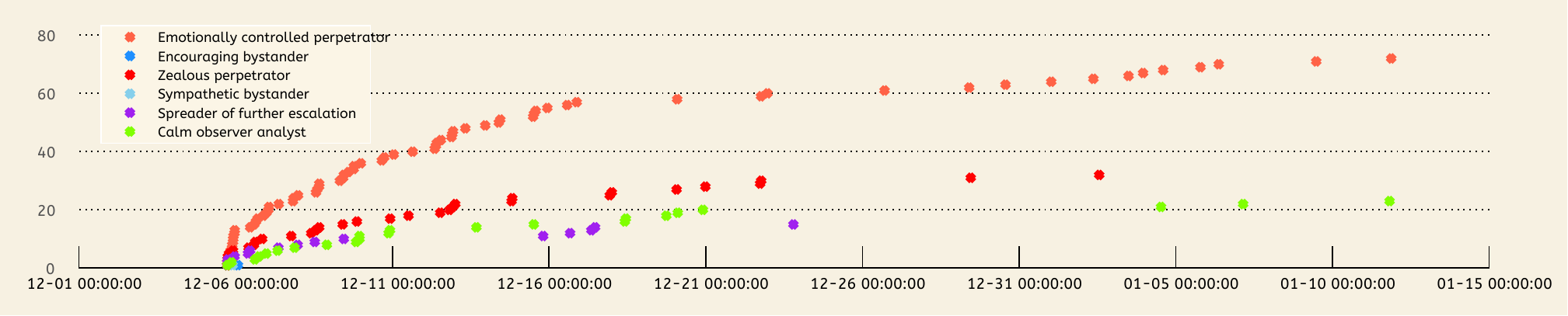}%
\label{Fig4e}}
\caption{\centering{Detail of the shaded section in Fig. \protect\ref{Fig2}\protect\subref{Fig2f} - \protect\ref{Fig2}\protect\subref{Fig2j}. (a) Scenario 6, (b) Scenario 7, (c) Scenario 8, (d) Scenario 9, (e) Scenario 10. The abscissa represents time while the ordinate represents the number of people occupied in different roles.}}
\label{Fig4}
\end{figure*}

\textbf{3) The Shengbin Lin Case (Scenario 3)}

In this scenario, the main character, Lin, used to present himself as a victim. However, with the recent remarriage, the birth of his child, the previous infidelity in marriage, cheating on women’s feelings, exciting public opinion, etc., the direction of public opinion has been reversed. As the situation evolved, the most significant proportion was exaggerated and fueled bystanders, who tried to steer the conversation unfavourably. Since there was no conclusive evidence or official involvement, calm observer analysts make up the second largest percentage from Fig. 1.

Fig. 2(c) shows a small number of members from July to August 2021. During this period, Lin announced the news of his remarriage and childbirth on Weibo. Then, he was suspected of infidelity in marriage by netizens, which led to the attack on public opinion. At this stage, most members were occupied with zealous and emotionally controlled perpetrators. However, public opinion soon subsided, and a new wave of public opinion emerged in early 2022. Fig. 3(c) embodies the shadows of Fig. 2(c). We can see that the most numerous roles are from exaggerated and fueled bystanders, followed by calm observer analysts. The main reason for this phenomenon is that despite the overwhelming skepticism, no police intervention was involved, and there was no hard evidence. Therefore, netizens with the mentality of “eating melon” were more likely to promote the incident by embellishing it and expecting to expand it further. There was also a growing group of calm observer analysts, who were largely neutral and waited to see what happened.

\textbf{4) The Jinglei Lee Divorce Case (Scenario 4)}

In Fig. 1, the proportion of calm observer analysts is highest due to the absence of solid evidence from Lee when the accusation on Weibo provided by Lee has set off the heat of public opinion. Most people still analyzed calmly as a bystander without excessive remarks. However, due to the massive influence of Wang Leehom in the Chinese music world, his fans would turn to Lee after their idol was attacked and questioned, so there were a considerable proportion of zealous perpetrators and emotionally controlled perpetrators.

As seen from Fig. 2(d), public opinion started at the end of February 2022, when the divorce dispute between the two people started, thus attracting more public attention. We focus on the analysis from the end of February to the middle of March 2022, as shown in Fig. 3(d). We can see that spreaders of further escalation have always occupied a high proportion, which is mainly because Wang is a very well-known figure who attracted the attention of many influencers and entertainment records. At the same time, there was also a phenomenon of using Wang’s popularity to gain trending topics. In addition, the number of emotionally controlled perpetrators to Lee was initially ahead of calm observer analysts, especially on March 5, when the number of emotionally controlled perpetrators increased rapidly. However, after March 10, the numbers were close to the same. This is because the dispute over visitation rights in the divorce case was heard in New York on March 4, and Wang has not seen the children for more than four months. Wang has applied for visitation many times before, but Lee Jinglei refused. On the contrary, Lee was still platonic and did not submit new evidence. This prompted a fair number of netizens sympathetic to Wang to condemn Lee. After the March 5 court hearing, Wang’s wish to see his children finally became possible, according to an official filing of the case. This led to a slower growth rate in the number of emotionally controlled perpetrators to Lee and more calm observer analysts who made more impartial and dispassionate judgments based on the latest news from the court.

\textbf{5) The Zheng Yu’s Cat Case (Scenario 5)}

In this scenario, due to the cruelty and visual impact of the injured cat in the drama, it suffered attacks from many cat-loving Weibo users, with more than half of the members being zealous perpetrators and emotionally controlled perpetrators from Fig. 1. The crew clarified based on this. However, it was not convincing enough to be questioned by most netizens. In addition, the proportion of encouraging bystanders was small.

As shown in Fig. 2(e), the crew’s Weibo account received mixed comments after the incident, with the number of comments rising rapidly. We focus on Fig. 3(e), the shaded part of Fig. 2(e). From Fig. 3(e), the number of members occupied by zealous perpetrators, emotionally controlled perpetrators, and exaggerated and fueled bystanders increased rapidly after the incident. This is because the young people who pay attention to this incident are primarily netizens who like cats. At the same time, based on the initial statements from the cast, there were a fair amount of members in calm observer analysts.

\textbf{6) The Ge Jiang’s Mother Case (Scenario 6)}

This incident began when Chen Lan, a writer, complained that Jiang Ge’s mother should disclose the details of the donations she received, and Jiang Ge’s mother sued Chen for slander for a long time. Jiang Ge’s case has been concluded before, and Jiang Ge’s mother won the lawsuit and has been widely sympathized with and supported by society as a shidu mother. As a result, a more significant proportion of members were sympathetic bystanders, calm observer analysts, and encouraging bystanders. However, since then, Jiang Ge’s mother has applied for the trademark registration of Jiang Ge, which has been controversial among many netizens. The controversy on the Internet mainly included emotionally controlled perpetrators who expressed their dissatisfaction with this incident and spreaders of further escalation who used their influence to try to cause a more significant storm of public opinion.

From Fig. 2(f), after public opinion began, the number of members represented by different roles increased rapidly in a short time. We chose the shaded part of Fig. 2(f) to analyze the role changes in detail. As shown in Fig. 4(a), different from the previous scenarios, the number of sympathetic bystanders and encouraging bystanders was significantly higher, and the number of zealous perpetrators was lower, which reflected the public’s sympathy and care for this shidu mother. However, the rapid increase in the number of emotionally controlled perpetrators also shows that netizens questioned the controversial actions of Jiang Ge’s mother. In terms of growth rate, in the early hours of March 13, the growth rate of members slowed down, but during the day, the growth rate started to pick up again.

\textbf{7) The Account Selling Case (Scenario 7)}

The incident occurred after the China Eastern Airlines plane crash that killed everyone on board. People expressed their deep condolences for the dead passengers on the plane, but a girl selling her ex-boyfriend’s dead game account caused outrage among many netizens. This behavior seriously violated moral ethics and did not consider the feelings of compatriots and relatives, so emotionally controlled perpetrators and zealous perpetrators accounted for more than 70\% from Fig. 1.

Fig. 2(g) shows that the number of zealous and emotionally controlled perpetrators was the largest, and the growth rate was fast. In addition, the number of perpetrators with a purpose who focus on games was also high. From the concrete analysis in Fig. 4(b), with time fermentation, members representing several types of roles were more concentrated in the figure, especially the emotionally controlled perpetrators. This shows that the majority of netizens have zero tolerance for behavior that challenges the moral bottom line. At the same time, spreaders of further escalation also accelerated the spread of public opinion.

\textbf{8) The Confirmed COVID-19 Patient Case (Scenario 8)}

In this scenario, the daily life of many people is affected as the girl is confirmed to be COVID-19 positive, and the epidemiological examination shows her going to multiple public places. As a result, there was a high proportion of zealous perpetrators and emotionally controlled perpetrators, who were usually motivated by the unwanted influence of the girls’ messages on their lives. In addition, since the diagnosis was not due to false reporting and the girl was not subjectively at fault, sympathetic bystanders and encouraging bystanders gave more sympathy and encouragement to the victim.

In Fig. 2(h), there are more roles with positive emotions, such as sympathetic bystanders and encouraging bystanders, than in other scenarios. Since the data only spanned three days, we use a more detailed time chart analysis. As shown in Fig. 4(c), the number of members with different roles rose rapidly after the incident began. Under the role of spreaders of further escalation, the growth rate was relatively fast. However, sympathetic bystanders and encouraging bystanders were outnumbered most of the time by zealous perpetrators, emotionally controlled perpetrators, and exaggerated and fueled bystanders. At the beginning of the episode, the rate and number of zealous perpetrators increased to be similar to or even higher than that of sympathetic bystanders and encouraging bystanders. This is because most of the members in such roles were affected by the girl’s flow, which disturbed their daily lives, resulting in a relatively intense and rapid outbreak of dissatisfaction. With a higher proportion of positive roles, the whole story has become more positive, thus cleansing the scene of negative voices to the greatest extent.

\textbf{9) The Jing Wu’s IP Address Case (Scenario 9)}

Wu Jing has acted in several patriotic films in recent years and has been branded a patriot himself. To our surprise, his IP was displayed overseas in the current scenario. Due to the frequent occurrence of COVID-19 in China recently, netizens questioned whether he went abroad to evade the control of the epidemic. Hence, the proportion of zealous perpetrators and exaggerated and fueled bystanders is high, as seen in Fig. 1. However, as the incident was clarified, calm observer analysts and encouraging bystanders looked at the incident more rationally rather than through gratuitous verbal attacks.

As shown in Fig. 2(i), the incident lasted for a long time, with the largest number of zealous perpetrators and exaggerated and fueled bystanders. We select the period from April 21 to May 3 to focus on analysis, as shown in Fig. 4(d). The cause of the incident was April 30, when a netizen found that Wu’s IP territory was Thailand. Therefore, before April 30, the number of members represented by different roles was relatively balanced, the number of roles with negative emotions was also small, and the overall picture was harmonious. After April 30, however, as public opinion grew, the number of zealous perpetrators and exaggerated and fueled bystanders increased sharply, surpassing the number of calm observer analysts and encouraging bystanders. At this point, the overall atmosphere of public opinion under the scene became increasingly disharmonious.

\textbf{10) The Godfrey Gao’s Death Case (Scenario 10)}

In this scenario, Godfrey Gao suffered sudden cardiac death during the early morning recording of Zhejiang TV’s Chase Me. Because Gao is a famous star among fans, many fans pointed the finger at the Zhejiang Satellite TV’s Chase Me team, where zealous perpetrators and emotionally controlled perpetrators accounted for more than 70\% of the members.

From Fig. 2(j), the number of zealous and emotionally controlled perpetrators has exploded since the beginning of the incident. We selected this stage for detailed analysis, as shown in Fig. 4(e). In Fig. 4(e), the scope of public opinion constantly expanded due to Gao’s enormous influence and the increasing number of spreaders of further escalation. Many of Gao’s fans launched a long and extensive verbal attack on Zhejiang TV. During this period, there were a few calm observer analysts and encouraging bystanders, but the public opinion atmosphere could not be changed as a whole.

\subsection{Factors influencing the distribution of roles}
\noindent
Based on the nine typical roles under ten cyberbullying scenarios, we sum up their similarities and differences and preliminarily obtain the factors that affect the distribution of roles of the edge crowd.

\subsubsection{Topic type \& gender}
\noindent
In previous studies, many scholars have paid attention to the issues of the relationship between gender and cyberbullying. For example, Olweus et al.\upcite{80} argued that males and females engage in traditional in-person bullying to different degrees, with males more likely to commit bullying. Kowalski et al.\upcite{81} found that girls are more likely to be victims of cyberbullying than boys.

In our work, we find that the topic type \& gender are factors influencing the distribution of an occupied number of roles involved in cyberbullying. Concretely, Scenario 1 consists of the topic of sports, which is male-dominated. Therefore, the proportion of males in roles in Scenario 1 is much higher than that of females. Scenario 4 and Scenario 10 involve topics about entertainment that attract more females. Hence, the proportion of females under these two scenarios is higher than that of males. In many topics, there is no clear gender difference, such as in Scenario 8, where we cannot find a clear difference. 

\subsubsection{Official intervention}
\noindent
In response to cyberbullying, many countries or institutions have enacted legislation, and a series of policies or measures formulated by the authorities can effectively reduce the harm of cyberbullying. In 2019, Tiiri et al.\upcite{82} explored the impact of introducing a National Anti-bullying program, KiVa, in Finnish schools in 2009. This school-wide intervention treats bullying as a group phenomenon and works to reduce bystanders, thereby reducing the motivation for bullies and their support for bullying. Legislation in most Gulf states takes the form of general cybercrime laws, which aim to punish various types of cybercrime\upcite{83}. In Canada, through a range of legal and policy responses, the broad range of behaviors and impacts associated with cyberbullying or online hate spreading can be addressed\upcite{84}.

In this work, whether the official department intervenes or gives the truth significantly impacts the distribution of roles. In Scenario 2, the police intervened and finally provided the case’s progress so that the perpetrators were no longer one-sided. In Scenario 4, the number of members in sympathetic and negative roles was not significantly different initially. Still, after March 5, the difference between the number of sympathetic members and those in negative roles gradually grew from Fig. 3(d). Part of the reason is that while Wang submitted eleven documents as evidence to the court, Lee had not presented sufficient evidence, leading to criticism about her credibility. Therefore, the number of Lee’s sympathetic roles grew slower than the number of negative ones. For comparison, in Scenario 5, the government did not intervene initially. As a result, the number of zealous perpetrators and emotionally controlled perpetrators keeps increasing. There was no official involvement in Scenario 10, and the evolution trend of the number of zealous and emotionally controlled perpetrators did not change significantly.

\subsubsection{Ethics \& morality}
\noindent
In Ref. \cite{85}, moral outrage is a comprehensive reaction of cognition, emotion and behavior, including indignation against injustice and helping vulnerable groups restore fairness, mainly triggered by witnessing unfair incidents. Hoffman et al.\upcite{86} also defined moral outrage as the anger aroused when moral norms are violated. In most focus incidents that form a particular scale of network collective action, some netizens always have angry emotions about some incidents that violate normal norms or go beyond the bottom line of psychological acceptance. These angry emotions are mostly built on some incidents that violate the bottom line of people’s emotional values, such as economic corruption, private gain, and moral bankruptcy\upcite{87}. In this paper, challenging moral boundaries is analogous to moral outrage.

In this work, whether or not to challenge the ethical bottom line of the public is also an essential factor influencing the distribution of the occupied number of roles. In Scenario 7, the MU5375 crash has brought Chinese people into grief, but the girl selling the game account of her dead ex-boyfriend for profit challenged the bottom line of human ethics. In this scenario, the proportion of zealous perpetrators and emotionally controlled perpetrators totalled 70.75\%, the highest of all the scenarios.

\subsubsection{Victim’s subjective fault}
\noindent
The subjective experience induced when individuals evaluate their own or others’ actions according to moral standards and social conditions is called moral emotions, such as guilt, pride, shame, empathy and other emotions\upcite{88}. The public will protest and express moral sentiments when they observe others acting contrary to usual standards, unjust or unfair, and committing any other moral or social transgression\upcite{89}.

In this work, the victim’s subjective fault also influences the number of people in different roles. In Scenario 7, the victim maliciously stole an MU5375 passenger’s account for profit. This behavior belongs to the subjective fault. Netizens abused this victim with subjective faults in a direct way. Therefore, it can be seen that the number of members in negative roles, such as zealous perpetrators and emotionally controlled perpetrators, grew strongly and maintained a high proportion in Figs. 2(g) and 4(b).

\subsubsection{Historical public perception of victims}
\noindent
In the case presented in this paper, the victim’s historical location and behavior are similar to a widespread phenomenon, namely the “collapse of human structure”. The “collapse of human structure” refers to destroying its image due to something that overturns the original positive impression left to everyone before\upcite{90}. The media image in communication studies corresponds to the concept of human design, which refers to the image presented in the media. However, due to commercial interests, entertainment effects and other factors, the final image given online differs greatly from what people do and say in real life, and human design will collapse overnight.

In this work, we find that the public image of the victim before cyberbullying also impacts the distribution of an occupied number of roles. In Scenario 3, Lin was previously presented as an arson victim. However, the news of his remarriage led to speculation of infidelity, which is a far cry from his previous image. Thus, members in negative roles, such as zealous perpetrators, were more numerous. The same case also includes Wu in Scenario 9. He had always presented himself as a patriotic actor, so his overseas IP address led to groundless speculation and slander. Different from Scenario 3 and Scenario 9, in Scenario 6, Ge Jiang’s mother had been suffering from the loss of her only daughter. Hence, most netizens still sympathize and encourage Jiang’s mother despite the doubts.

In summary, based on the modeling method for role features proposed in this paper, we propose a clustering algorithm named DEK in the mixed categorical-continuous space. Experiments on different datasets verify the superiority and effectiveness of our proposed role identification method. Finally, we conduct fine-grained and multi-level role identification under different cyberbullying scenarios. We put the identified roles into the original cyberbullying scenarios for analysis and verification. The conclusions obtained from the analysis and verification are consistent with common sense knowledge of real cyberbullying scenarios, and relevant previous qualitative analysis studies can support our analysis results.

\section{Conclusion}
\noindent
In this paper, to obtain deeper insights and differentiate roles for cyberbullying in edge AI applications on social edge computing devices, we designed algorithms that run directly on mobile devices, enabling each mobile device to detect the roles involved in cyberbullying. Our proposed DEK considered the problem brought by mixed variables in the actual cyberbullying scenarios and the problem that it is easy to fall into the local optimal during the clustering process. DEK can handle the fine-grained division of roles in cyberbullying and is tested on five public datasets, proving our algorithm’s universality. In the empirical analysis, we obtained nine types of cyberbullying roles and analyzed the relevant factors affecting the number of occupied roles. Our work in this paper can be placed in devices at the edge of the Internet, leading to better real-time role identification performance for cyberbullying.

The fine-grained role-playing behavior pattern obtained will help future research to have a more profound insight into the nature and propagation rules of cyberbullying to conduct more effective interventions against cyberbullying. This work has some limitations. Introducing Gower distance increases the calculation time, increasing the algorithm’s parameter sensitivity. Therefore, it is necessary to improve the efficiency and robustness of the algorithm in the future. In addition, we collected real data for each cyberbullying scenario, but the proportion of this data to the total data for the current scenario is unknown. As a result, the number of role types we found using DEK may be less than or equal to the number of actual roles present in real cyberbullying scenarios. In this paper, we still identified the types of cyberbullying roles through manual interpretation to analyze the clusterings. There may be some subjective deviation. In the future, it is necessary to extend this method to more platforms to prevent cyberbullying. Further, developing more DEK-assisted AI products for social edge computing will strengthen the governance of cyberbullying.

\section*{Acknowledgements}
This work was supported by the National Key Research and Development Program of China under Grant No. 2021YFC3300202. Tun Lu is also a faculty of Shanghai Key Laboratory of Data Science, Fudan Institute on Aging, MOE Laboratory for National Development and Intelligent Governance, and Shanghai Institute of Intelligent Electronics \& Systems, Fudan University.

\renewcommand\refname{\large

\end{document}